\documentclass{elsart}
\usepackage{graphicx,amssymb,amsfonts,amsmath}

\newcommand{\mcm}[3]{\newcommand{#1}[#2]{{\ensuremath{#3}}}}

\usepackage{latexsym}
\usepackage{amssymb}

\mcm{\blank}{0}{(\emptybk)} \mcm{\dashbk}{0}{\mbox{---}}
\mcm{\emptybk}{0}{\:\:} \mcm{\hyph}{0}{\mbox{-}}

\mcm{\diagspace}{0}{\mbox{\hspace{2em}}}

\mcm{\cat}{1}{\mc{#1}} \mcm{\fcat}{1}{\mb{#1}}
\mcm{\mc}{1}{\mathcal{#1}} \mcm{\mr}{1}{\mathrm{#1}}
\mcm{\mi}{1}{\mathit{#1}} \mcm{\mb}{1}{\mathbf{#1}}
\mcm{\scat}{1}{\Bbb{#1}} \mcm{\twid}{1}{\widetilde{#1}}

\mcm{\elt}{0}{\in} \mcm{\sub}{0}{\,\subseteq\,}
\mcm{\such}{0}{\:|\:} \mcm{\without}{0}{\setminus}

\mcm{\atsr}{0}{\Box} \mcm{\eqv}{0}{\,\simeq\,}
\mcm{\iso}{0}{\,\cong\,}
\mcm{\of}{0}{\raisebox{0.2mm}{\ensuremath{\scriptstyle\circ}}}

\mcm{\bdry}{0}{\partial}

\mcm{\Bee}{0}{\cat{B}} \mcm{\Beep}{0}{\cat{B'}}
\mcm{\Eee}{0}{\cat{E}} \mcm{\Eeep}{0}{\cat{E'}}

\mcm{\Ess}{0}{\cat{S}} \mcm{\Tee}{0}{\cat{T}}
\mcm{\Teep}{0}{\cat{T'}} \mcm{\Stee}{0}{\scat{T}}
\mcm{\Steep}{0}{\scat{T'}}

\mcm{\blbk}{0}{\blank^{\blob}}
\mcm{\blob}{0}{\scriptscriptstyle{\bullet}}
\mcm{\stbk}{0}{\blank^{*}} \mcm{\ubl}{0}{{}^{\blob}}
\mcm{\ust}{0}{{}^{*}}

\mcm{\Cartpr}{0}{\pr{\Eee}{T}} \mcm{\Cartprp}{0}{\pr{\Eeep}{T'}}
\mcm{\Mnd}{0}{\triple{T}{\eta}{\mu}}
\mcm{\Zeropr}{0}{\pr{\Set}{\id}}

\mcm{\dopset}{0}{\ftrcat{\Delta^{\op}}{\Set}}
\mcm{\tropset}{0}{\ftrcat{\fcat{TR}^{\op}}{\Set}}

\mcm{\cod}{0}{\mr{cod}} \mcm{\dom}{0}{\mr{dom}}
\mcm{\End}{0}{\mr{End}} \mcm{\Hom}{0}{\mr{Hom}}
\mcm{\ob}{0}{\mr{ob}\,} \mcm{\op}{0}{\mr{op}}

\mcm{\comp}{0}{\mi{comp}} \mcm{\id}{0}{\mi{id}}
\mcm{\ids}{0}{\mi{ids}} \mcm{\mult}{0}{\mi{mult}}
\mcm{\unit}{0}{\mi{unit}}

\mcm{\Ab}{0}{\fcat{Ab}} \mcm{\Alg}{0}{\fcat{Alg}}
\mcm{\Bim}{1}{\fcat{Bim}(#1)} \mcm{\Cat}{0}{\fcat{Cat}}
\mcm{\Cay}{0}{\fcat{Cay}} \mcm{\Cpn}{1}{\pr{\Set/S_{#1}}{T_{#1}}}
\mcm{\fc}{0}{\fcat{fc}} \mcm{\fm}{0}{\fcat{fm}}
\mcm{\Graph}{0}{\fcat{Graph}} \mcm{\Gy}{0}{\fcat{Gy}}
\mcm{\Hpn}{1}{\pr{\Eee_{#1}}{P_{#1}}} \mcm{\Mon}{0}{\mb{Mon}}
\mcm{\Multicat}{0}{\fcat{Multicat}} \mcm{\One}{0}{\fcat{1}}
\mcm{\PD}{1}{\fcat{PD}_{#1}} \mcm{\Prof}{0}{\fcat{Prof}}
\mcm{\Set}{0}{\fcat{Set}} \mcm{\Span}{0}{\fcat{Span}}
\mcm{\Ssq}{0}{\fcat{Ssq}} \mcm{\Struc}{0}{\fcat{Struc}}
\mcm{\Sym}{0}{\fcat{Sym}} \mcm{\TR}{1}{\fcat{TR}(#1)}
\mcm{\Tr}{0}{\fcat{Tr}} \mcm{\Twocat}{0}{\fcat{2\hyph\Cat}}

\mcm{\integers}{0}{\mathbb{Z}}

\mcm{\range}{2}{#1,\,\ldots\,,#2}
\mcm{\bftuple}{2}{\tuplebts{\range{#1}{#2}}}
\mcm{\tuple}{3}{\tuplebts{\range{#1,#2}{#3}}}
\mcm{\rttuple}{1}{\tuplebts{\,\ldots\,,#1}}
\mcm{\abftuple}{2}{\atuplebts{\range{#1}{#2}}}
\mcm{\atuple}{3}{\atuplebts{\range{#1,#2}{#3}}}
\mcm{\arttuple}{1}{\atuplebts{\,\ldots\,,#1}}
\mcm{\sqbftuple}{2}{\obt\range{#1}{#2}\cbt}
\mcm{\pr}{2}{\tuplebts{#1,#2}}
\mcm{\triple}{3}{\tuplebts{#1,#2,#3}}

\mcm{\eend}{2}{#1[#2]} \mcm{\ehom}{3}{#1[#2,#3]}
\mcm{\ftrcat}{2}{[#1,#2]} \mcm{\homset}{3}{#1(#2,#3)}
\mcm{\multihom}{3}{#1(#2;#3)}
\mcm{\relhom}{5}{#1_{#2}(\range{#3}{#4};#5)}

\mcm{\go}{0}{\rTo} \mcm{\goby}{1}{\rTo^{#1}}
\mcm{\goesto}{0}{\,\longmapsto\,} \mcm{\goiso}{0}{\goby{\diso}}
\mcm{\monic}{0}{\rMonic} \mcm{\og}{0}{\lTo}
\mcm{\ogby}{1}{\lTo^{#1}}

\mcm{\gph}{2}{\spn{#1}{T #2}{#2}} \mcm{\graph}{4}{\spaan{#1}{T
#2}{#2}{#3}{#4}} \mcm{\oppair}{2}{\stackrel{\rTo^{#1}}{\lTo_{#2}}}
\mcm{\parpair}{2}{\stackrel{\rTo^{#1}}{\rTo_{#2}}}
\mcm{\spn}{3}{#2 \og #1 \go #3} \mcm{\spaan}{5}{#2 \ogby{#4} #1
\goby{#5} #3}

\mcm{\bktdvslob}{3}
    {\left(
    \begin{diagram}[height=1.5em]
    #1      \\
    \dTo>{\,#2} \\
    #3      \\
    \end{diagram}
    \right)}
\mcm{\slob}{3}{(#1 \goby{#2} #3)} \mcm{\vslob}{3}
    {\left.
    \begin{diagram}[height=1.5em]
    #1      \\
    \dTo>{\,#2} \\
    #3      \\
    \end{diagram}
    \right.}

\newenvironment{tree}
    {\begin{diagram}[height=1em,width=.75em,abut,noPS,tight]}
    {\end{diagram}}

\mcm{\enode}{0}{\circ}

\mcm{\nl}{1}{\stackrel{\textstyle #1}{\node}}
\mcm{\node}{0}{\bullet}

\mcm{\utree}{0}{\node}

\mcm{\diso}{0}{\sim}

\mcm{\vdiso}{0}{\wr}

\mcm{\nat}{0}{\mathbb{N}}

\mcm{\Onepr}{0}{\pr{\Graph}{\fc}}
\newlength{\nllwidth}
\newlength{\nllheight}
\newcommand{\stackbelow}[2]{%
\settowidth{\nllwidth}{\ensuremath{#1}\ensuremath{#2}}%
\settoheight{\nllheight}{\ensuremath{#2}}%
\addtolength{\nllheight}{.3ex}%
\mbox{%
\ensuremath{#1}%
\hspace{-.5\nllwidth}%
\raisebox{-1\nllheight}{\ensuremath{#2}}}}
\mcm{\nlal}{2}{\stackbelow{\nl{#1}}{#2}}
\mcm{\nll}{1}{\stackbelow{\node}{#1}} \mcm{\wun}{0}{\fcat{1}}
\mcm{\atuplebts}{1}{\langle #1 \rangle} \mcm{\tuplebts}{1}{(#1)}
\mcm{\bo}{0}{(} \mcm{\bc}{0}{)}
\mcm{\UBilax}{0}{\fcat{UBicat}_\mr{lax}}
\mcm{\UBiwk}{0}{\fcat{UBicat}_\mr{wk}}
\mcm{\UBistr}{0}{\fcat{UBicat}_\mr{str}}
\mcm{\Bilax}{0}{\fcat{Bicat}_\mr{lax}}
\mcm{\Biwk}{0}{\fcat{Bicat}_\mr{wk}}
\mcm{\Bistr}{0}{\fcat{Bicat}_\mr{str}} \mcm{\rotsub}{0}{\cup
\raisebox{0.1em}{$\scriptstyle{|}$}} \mcm{\pd}{0}{\fcat{pd}}
\mcm{\rep}{1}{\widehat{#1}} \mcm{\ovln}{1}{\overline{#1}}
\mcm{\Gph}{0}{\fcat{Gph}} \mcm{\tr}{0}{\fcat{tr}}

\mcm{\ladj}{0}{\,\dashv\,} \mcm{\zeropd}{0}{\node}
    {\end{diagram}}
\mcm{\END}{0}{\fcat{End}} \mcm{\HOM}{0}{\fcat{Hom}}

\newlength{\gwidth} 
\newlength{\gvert}  
\newlength{\gdrop}  
\newlength{\gbaredrop}  
\newlength{\goffset}    
\newlength{\gtemp}  

\newcommand{\present}[1]{%
\makebox[1\gwidth]{%
\rule[-1\gdrop]{0ex}{1\gvert}%
\raisebox{-1\gbaredrop}{#1}}}

\newcommand{\presentl}[1]{%
\makebox[1\gwidth][l]{%
\rule[-1\gdrop]{0ex}{1\gvert}%
\raisebox{-1\gbaredrop}{#1}}}

\newcommand{\presentr}[1]{%
\makebox[1\gwidth][r]{%
\rule[-1\gdrop]{0ex}{1\gvert}%
\raisebox{-1\gbaredrop}{#1}}}

\newcommand{\ginitdims}[2]{
\setlength{\unitlength}{1em}
\setlength{\goffset}{.25\unitlength}
\setlength{\gwidth}{#1\unitlength}
\setlength{\gvert}{#2\unitlength}
\setlength{\gdrop}{.5\gvert}
\addtolength{\gdrop}{-1\goffset}
\setlength{\gbaredrop}{1\gdrop}
\addtolength{\gvert}{.6\unitlength}
\addtolength{\gdrop}{.3\unitlength}}    

\newcommand{\cinitdims}[2]{
\setlength{\unitlength}{1em}
\setlength{\goffset}{.35\unitlength}
\setlength{\gwidth}{#1\unitlength}
\setlength{\gvert}{#2\unitlength}
\setlength{\gdrop}{.5\gvert}
\addtolength{\gdrop}{-1\goffset}
\setlength{\gbaredrop}{1\gdrop}
\addtolength{\gvert}{.6\unitlength}
\addtolength{\gdrop}{.3\unitlength}}    

\newcommand{\gsinitdims}[2]{
\setlength{\unitlength}{0.5em}
\setlength{\goffset}{.25\unitlength}
\setlength{\gwidth}{#1\unitlength}
\setlength{\gvert}{#2\unitlength}
\setlength{\gdrop}{.5\gvert}
\addtolength{\gdrop}{-1\goffset}
\setlength{\gbaredrop}{1\gdrop}
\addtolength{\gvert}{.6\unitlength}
\addtolength{\gdrop}{.3\unitlength}}    

\newcommand{\sidespic}[1]{%
\settowidth{\gtemp}{\ensuremath{#1}}%
\addtolength{\gwidth}{1\gtemp}}

\newcommand{\abovepic}[1]{%
\settoheight{\gtemp}{\ensuremath{#1}}%
\addtolength{\gvert}{1\gtemp}%
\settodepth{\gtemp}{\ensuremath{#1}}%
\addtolength{\gvert}{1\gtemp}}

\newcommand{\belowpic}[1]{%
\settoheight{\gtemp}{\ensuremath{#1}}%
\addtolength{\gvert}{1\gtemp}%
\addtolength{\gdrop}{1\gtemp}%
\settodepth{\gtemp}{\ensuremath{#1}}%
\addtolength{\gvert}{1\gtemp}%
\addtolength{\gdrop}{1\gtemp}}

\newcommand{\cell}[4]{\put(#1,#2){\makebox(0,0)[#3]{\ensuremath{#4}}}}
\mcm{\zmark}{0}{\scriptstyle{\bullet}}

\newcommand{\pregfst}[1]{%
\begin{picture}(0.5,0.2)(-0.5,-0.2)%
\cell{-0.1}{-0.2}{tr}{#1}%
\cell{0}{0}{c}{\zmark}%
\end{picture}}

\mcm{\gfst}{1}{%
\ginitdims{0.5}{0.4}%
\sidespic{#1}%
\belowpic{#1}%
\presentr{\pregfst{#1}}}

\newcommand{\preglst}[1]{%
\begin{picture}(0.5,0.2)(0,-0.2)%
\cell{0.1}{-0.2}{tl}{#1}%
\cell{0.05}{0}{c}{\zmark}%
\end{picture}}

\mcm{\glst}{1}{%
\ginitdims{.5}{.4}%
\sidespic{#1}%
\belowpic{#1}%
\presentl{\preglst{#1}}}

\newcommand{\preglft}[1]{%
\begin{picture}(0,0.2)(0,-0.2)%
\cell{-0.1}{-0.2}{tr}{#1}%
\cell{0.05}{0}{c}{\zmark}%
\end{picture}}

\mcm{\glft}{1}{%
\ginitdims{0}{.4}%
\belowpic{#1}%
\present{\preglft{#1}}}

\newcommand{\pregrgt}[1]{%
\begin{picture}(0,0.2)(0,-0.2)%
\cell{0.1}{-0.2}{tl}{#1}%
\cell{0.05}{0}{c}{\zmark}%
\end{picture}}

\mcm{\grgt}{1}{%
\ginitdims{0}{.4}%
\belowpic{#1}%
\present{\pregrgt{#1}}}

\newcommand{\pregblw}[1]{%
\begin{picture}(0,0.3)(0,-0.3)
\cell{0}{-0.3}{t}{#1}%
\cell{0.05}{0}{c}{\zmark}%
\end{picture}}

\mcm{\gblw}{1}{%
\ginitdims{0}{.6}%
\belowpic{#1}%
\present{\pregblw{#1}}}

\newcommand{\pregfbw}[1]{%
\begin{picture}(0,0.65)(0,-0.65)
\cell{0}{-0.65}{t}{#1}%
\cell{0.05}{0}{c}{\zmark}%
\end{picture}}

\mcm{\gfbw}{1}{%
\ginitdims{0}{1.3}%
\belowpic{#1}%
\present{\pregfbw{#1}}}

\newcommand{\pregzero}[1]{%
\begin{picture}(0.8,0.4)(-0.4,-0.4)
\cell{0}{-0.4}{t}{#1}%
\cell{0}{0}{c}{\zmark}%
\end{picture}}

\mcm{\gzero}{1}{%
\ginitdims{0.8}{.6}%
\belowpic{#1}%
\sidespic{#1}%
\present{\pregzero{#1}}}

\newcommand{\pregone}[1]{%
\begin{picture}(5,0.4)(0,-0.2)%
\cell{2.5}{0.2}{b}{#1}%
\put(0,0){\vector(1,0){5}}%
\end{picture}}

\mcm{\gone}{1}{%
\ginitdims{5}{0.4}%
\abovepic{#1}%
\present{\pregone{#1}}}

\newcommand{\pregtwo}[3]{%
\begin{picture}(5,3.4)(0,-0.2)%
\cell{2.5}{3.2}{b}{#1}%
\cell{2.5}{-.2}{t}{#2}%
\cell{2.7}{1.5}{l}{#3}%
\qbezier(0,1.5)(2.5,4.5)(5,1.5)%
\qbezier(0,1.5)(2.5,-1.5)(5,1.5)%
\put(5,1.5){\vector(1,-1){0}}%
\put(5,1.5){\vector(1,1){0}}%
\put(2.5,2.5){\vector(0,-1){2}}%
\end{picture}}

\mcm{\gtwo}{3}{%
\ginitdims{5}{3.4}%
\abovepic{#1}%
\belowpic{#2}%
\present{\pregtwo{#1}{#2}{#3}}}

\newcommand{\pregthree}[5]{%
\begin{picture}(5,5.4)(0,-1.2)%
\cell{2.5}{4.2}{b}{#1}%
\cell{1.5}{1.7}{b}{#2}%
\cell{2.5}{-1.2}{t}{#3}%
\cell{2.7}{2.75}{l}{#4}%
\cell{2.7}{0.25}{l}{#5}%
\qbezier(0,1.5)(2.5,6.5)(5,1.5)%
\qbezier(0,1.5)(2.5,-3.5)(5,1.5)%
\put(0,1.5){\vector(1,0){5}}%
\put(2.5,3.5){\vector(0,-1){1.5}}%
\put(2.5,1){\vector(0,-1){1.5}}%
\put(5,1.5){\vector(1,-3){0}}%
\put(5,1.5){\vector(1,3){0}}%
\end{picture}}

\mcm{\gthree}{5}{%
\ginitdims{5}{5.4}%
\abovepic{#1}%
\belowpic{#3}%
\present{\pregthree{#1}{#2}{#3}{#4}{#5}}}

\newcommand{\pregfour}[7]{%
\begin{picture}(5,8.4)(0,-2.7)%
\cell{2.5}{5.7}{b}{#1}%
\cell{1.5}{2.8}{b}{#2}%
\cell{1.5}{0.2}{t}{#3}%
\cell{2.5}{-2.7}{t}{#4}%
\cell{2.7}{4.25}{l}{#5}%
\cell{2.7}{1.5}{l}{#6}%
\cell{2.7}{-1.25}{l}{#7}%
\qbezier(0,1.5)(2.5,9.5)(5,1.5)%
\qbezier(0,1.5)(2.5,4)(5,1.5)%
\qbezier(0,1.5)(2.5,-1)(5,1.5)%
\qbezier(0,1.5)(2.5,-6.5)(5,1.5)%
\put(2.5,5.25){\vector(0,-1){2}}%
\put(2.5,2.5){\vector(0,-1){2}}%
\put(2.5,-0.25){\vector(0,-1){2}}%
\put(5,1.5){\vector(1,-4){0}}%
\put(5,1.5){\vector(4,-3){0}}%
\put(5,1.5){\vector(4,3){0}}%
\put(5,1.5){\vector(1,4){0}}%
\end{picture}}

\mcm{\gfour}{7}{%
\ginitdims{5}{8.4}%
\abovepic{#1}%
\belowpic{#4}%
\present{\pregfour{#1}{#2}{#3}{#4}{#5}{#6}{#7}}}

\newcommand{\pregthreecell}[5]{%
\begin{picture}(8,5)(-4,-2.5)%
\cell{0}{2.5}{b}{#1}%
\cell{0}{-2.5}{t}{#2}%
\cell{-1.7}{0}{r}{#3}%
\cell{1.7}{0}{l}{#4}%
\cell{0}{0.2}{b}{#5}%
\qbezier(-4,0)(0,4.2)(4,0)%
\qbezier(-4,0)(0,-4.2)(4,0)%
\qbezier(-0.5,1.8)(-2.5,0)(-0.5,-1.8)%
\qbezier(0.5,1.8)(2.5,0)(0.5,-1.8)%
\put(-1,0){\vector(1,0){2}}%
\put(4,0){\vector(1,-1){0}}%
\put(4,0){\vector(1,1){0}}%
\put(-0.5,-1.8){\vector(1,-1){0}}%
\put(0.5,-1.8){\vector(-1,-1){0}}%
\end{picture}}

\mcm{\gthreecell}{5}{%
\ginitdims{8}{5}%
\abovepic{#1}%
\belowpic{#2}%
\present{\pregthreecell{#1}{#2}{#3}{#4}{#5}}}

\newcommand{\pregthreecellu}{%
\begin{picture}(5,3.4)(-0.5,-0.2)%
\qbezier(-.5,1.5)(2,4.5)(4.5,1.5)%
\qbezier(-.5,1.5)(2,-1.5)(4.5,1.5)%
\qbezier(1.5,2.7)(0.5,1.5)(1.5,0.3)%
\qbezier(2.5,2.7)(3.5,1.5)(2.5,0.3)%
\put(1.3,1.5){\vector(1,0){1.4}}%
\put(4.5,1.5){\vector(1,-1){0}}%
\put(4.5,1.5){\vector(1,1){0}}%
\put(1.5,0.3){\vector(2,-3){0}}%
\put(2.5,0.3){\vector(-2,-3){0}}%
\end{picture}}

\mcm{\gthreecellu}{0}{%
\ginitdims{5}{3.4}%
\present{\pregthreecellu}}

\newcommand{\pregtwocentre}[3]{%
\begin{picture}(5,3.4)(0,-0.2)%
\cell{2.5}{3.2}{b}{#1}%
\cell{2.5}{-.2}{t}{#2}%
\cell{2.5}{1.5}{c}{#3}%
\qbezier(0,1.5)(2.5,4.5)(5,1.5)%
\qbezier(0,1.5)(2.5,-1.5)(5,1.5)%
\put(5,1.5){\vector(1,-1){0}}%
\put(5,1.5){\vector(1,1){0}}%
\put(2.5,2.5){\vector(0,-1){2}}%
\end{picture}}

\mcm{\gtwocentre}{3}{%
\ginitdims{5}{3.4}%
\abovepic{#1}%
\belowpic{#2}%
\present{\pregtwocentre{#1}{#2}{#3}}}

\newcommand{\pregspecialone}[9]{%
\begin{picture}(8,8)(-4,-4)%
\cell{0}{3.9}{b}{#1}%
\cell{-2}{-0.2}{t}{#2}%
\cell{0}{-3.9}{t}{#3}%
\cell{-1.5}{1.1}{r}{#4}%
\cell{0.2}{1.5}{l}{#5}%
\cell{1.5}{1.1}{l}{#6}%
\cell{0.2}{-2}{l}{#7}%
\cell{-0.9}{2.3}{b}{#8}%
\cell{0.9}{2.3}{b}{#9}%
\qbezier(-4,0)(0,8)(4,0)%
\qbezier(-4,0)(0,-8)(4,0)%
\qbezier(-0.5,3.4)(-3.5,2)(-0.5,0.6)%
\qbezier(0.5,3.4)(3.5,2)(0.5,0.6)%
\put(-4,0){\vector(1,0){8}}%
\put(0,3.4){\vector(0,-1){2.8}}%
\put(0,-0.8){\vector(0,-1){2.4}}%
\put(-1.5,2.2){\vector(1,0){1.2}}%
\put(0.3,2.2){\vector(1,0){1.2}}%
\put(4,0){\vector(1,-2){0}}%
\put(4,0){\vector(1,2){0}}%
\put(-0.5,0.6){\vector(2,-1){0}}%
\put(0.5,0.6){\vector(-2,-1){0}}%
\end{picture}}

\mcm{\gspecialone}{9}{%
\ginitdims{8}{8}%
\abovepic{#1}%
\belowpic{#3}%
\present{\pregspecialone{#1}{#2}{#3}{#4}{#5}{#6}{#7}{#8}{#9}}}

\newcommand{\pregspecialtwo}{%
\begin{picture}(5,3.4)(0,-0.2)%
\qbezier(0,1.5)(2.5,4.5)(5,1.5)%
\qbezier(0,1.5)(2.5,-1.5)(5,1.5)%
\qbezier(1.7,2.5)(0,1.5)(1.7,0.5)%
\qbezier(3.3,2.5)(5,1.5)(3.3,0.5)%
\put(5,1.5){\vector(1,-1){0}}%
\put(5,1.5){\vector(1,1){0}}%
\put(1.7,0.5){\vector(3,-2){0}}%
\put(3.3,0.5){\vector(-3,-2){0}}%
\put(2.5,2.5){\vector(0,-1){2}}%
\put(1.2,1.5){\vector(1,0){1}}%
\put(2.8,1.5){\vector(1,0){1}}%
\end{picture}}

\mcm{\gspecialtwo}{0}{%
\ginitdims{5}{3.4}%
\present{\pregspecialtwo}}

\newcommand{\pregspecialthree}{%
\begin{picture}(5,5.4)(0,-1.2)%
\qbezier(0,1.5)(2.5,6.5)(5,1.5)%
\qbezier(0,1.5)(2.5,-3.5)(5,1.5)%
\qbezier(2,3.5)(1,2.75)(2,2)%
\qbezier(3,3.5)(4,2.75)(3,2)%
\qbezier(2,1)(1,0.25)(2,-0.5)%
\qbezier(3,1)(4,0.25)(3,-0.5)%
\put(0,1.5){\vector(1,0){5}}%
\put(1.5,2.75){\vector(1,0){2}}%
\put(1.5,0.25){\vector(1,0){2}}%
\put(5,1.5){\vector(1,-3){0}}%
\put(5,1.5){\vector(1,3){0}}%
\put(2,2){\vector(1,-1){0}}%
\put(3,2){\vector(-1,-1){0}}%
\put(2,-0.5){\vector(1,-1){0}}%
\put(3,-0.5){\vector(-1,-1){0}}%
\end{picture}}

\mcm{\gspecialthree}{0}{%
\ginitdims{5}{5.4}%
\present{\pregspecialthree}}

\newcommand{\pregonew}[1]{%
\begin{picture}(8,0.4)(0,-0.2)%
\cell{4}{0.2}{b}{#1}%
\put(0,0){\vector(1,0){8}}%
\end{picture}}

\mcm{\gonew}{1}{%
\ginitdims{8}{0.4}%
\abovepic{#1}%
\present{\pregonew{#1}}}

\mcm{\gzersu}{0}{%
\gsinitdims{0}{.6}%
\present{\pregblw{}}}

\mcm{\gonesu}{0}{%
\gsinitdims{5}{0.4}%
\present{\pregone{}}}

\mcm{\gtwosu}{0}{%
\gsinitdims{5}{3.4}%
\present{\pregtwo{}{}{}}}

\mcm{\gthreesu}{0}{%
\gsinitdims{5}{5.4}%
\present{\pregthree{}{}{}{}{}}}

\mcm{\gfoursu}{0}{%
\gsinitdims{5}{8.4}%
\present{\pregfour{}{}{}{}{}{}{}}}

\newcommand{\precone}[1]{%
\begin{picture}(4.2,0.4)(-0.3,-0.2)%
\cell{1.8}{0.2}{b}{#1}%
\put(0,0){\vector(1,0){3.6}}%
\end{picture}}

\mcm{\cone}{1}{%
\cinitdims{4.2}{0.4}%
\abovepic{#1}%
\present{\precone{#1}}}

\mcm{\gfstsu}{0}{%
\gsinitdims{0.5}{0.4}%
\presentr{\pregfst{}}}

\mcm{\glstsu}{0}{%
\gsinitdims{0.5}{0.4}%
\presentl{\preglst{}}}

\newcommand{\prectwodbl}[3]%
{\begin{picture}(4.2,3.4)(-0.1,-0.2)%
\cell{2}{3.2}{b}{#1}%
\cell{2}{-0.2}{t}{#2}%
\cell{2.3}{1.5}{l}{#3}%
\qbezier(0,2)(2,4)(4,2)%
\qbezier(0,1)(2,-1)(4,1)%
\put(4,2){\vector(1,-1){0}}%
\put(4,1){\vector(1,1){0}}%
\put(1.9,2.5){\line(0,-1){1.8}}%
\put(2.1,2.5){\line(0,-1){1.8}}%
\cell{2.01}{0.4}{b}{\vee}%
\end{picture}}

\mcm{\ctwodbl}{3}{%
\cinitdims{4.2}{3.4}%
\abovepic{#1}%
\belowpic{#2}%
\present{\prectwodbl{#1}{#2}{#3}}}

\newcommand{\precthreedbl}[5]{%
\begin{picture}(4.2,5.4)(-0.1,-0.2)%
\cell{2}{5.2}{b}{#1}%
\cell{1}{2.7}{b}{#2}%
\cell{2}{-.2}{t}{#3}%
\cell{2.3}{3.75}{l}{#4}%
\cell{2.3}{1.25}{l}{#5}%
\qbezier(0,3)(2,7)(4,3)%
\qbezier(0,2)(2,-2)(4,2)%
\put(0,2.5){\vector(1,0){4}}%
\put(1.9,4.5){\line(0,-1){1.3}}%
\put(2.1,4.5){\line(0,-1){1.3}}%
\cell{2.01}{2.9}{b}{\vee}%
\put(1.9,2){\line(0,-1){1.3}}%
\put(2.1,2){\line(0,-1){1.3}}%
\cell{2.01}{0.4}{b}{\vee}%
\put(4,3){\vector(1,-3){0}}%
\put(4,2){\vector(1,3){0}}%
\end{picture}}

\mcm{\cthreedbl}{5}{%
\cinitdims{4.2}{5.4}%
\abovepic{#1}%
\belowpic{#3}%
\present{\precthreedbl{#1}{#2}{#3}{#4}{#5}}}

\newcommand{\precthreecelltrp}[5]{%
\begin{picture}(8.2,5)(-4.1,-2.5)%
\cell{0}{2.5}{b}{#1}%
\cell{0}{-2.5}{t}{#2}%
\cell{-1.8}{0}{r}{#3}%
\cell{1.8}{0}{l}{#4}%
\cell{0}{0.3}{b}{#5}%
\qbezier(-4,0.5)(0,4)(4,0.5)%
\qbezier(-4,-0.5)(0,-4)(4,-0.5)%
\qbezier(-0.6,2)(-2.6,0)(-0.6,-2)%
\qbezier(-0.4,2)(-2.4,0)(-0.5,-1.9)%
\cell{-0.6}{-2}{b}{\lrcorner}%
\qbezier(0.4,2)(2.4,0)(0.5,-1.9)%
\qbezier(0.6,2)(2.6,0)(0.6,-2)%
\cell{0.65}{-2}{b}{\llcorner}%
\put(-1,0.15){\line(1,0){1.7}}%
\put(-1,0){\line(1,0){2}}%
\put(-1,-0.15){\line(1,0){1.7}}%
\cell{1.15}{0}{r}{>}%
\put(4,0.5){\vector(1,-1){0}}%
\put(4,-0.5){\vector(1,1){0}}%
\end{picture}}

\mcm{\cthreecelltrp}{5}{%
\cinitdims{8.2}{5}%
\abovepic{#1}%
\belowpic{#2}%
\present{\precthreecelltrp{#1}{#2}{#3}{#4}{#5}}}

\newcommand{\prectwo}[3]%
{\begin{picture}(4.2,3.4)(-0.1,-0.2)%
\cell{2}{3.2}{b}{#1}%
\cell{2}{-0.2}{t}{#2}%
\cell{2.2}{1.5}{l}{#3}%
\qbezier(0,2)(2,4)(4,2)%
\qbezier(0,1)(2,-1)(4,1)%
\put(4,2){\vector(1,-1){0}}%
\put(4,1){\vector(1,1){0}}%
\put(2,2.5){\vector(0,-1){2}}%
\end{picture}}

\mcm{\ctwo}{3}{%
\cinitdims{4.2}{3.4}%
\abovepic{#1}%
\belowpic{#2}%
\present{\prectwo{#1}{#2}{#3}}}

\newcommand{\precthree}[5]{%
\begin{picture}(4.2,5.4)(-0.1,-0.2)%
\cell{2}{5.2}{b}{#1}%
\cell{1}{2.7}{b}{#2}%
\cell{2}{-.2}{t}{#3}%
\cell{2.2}{3.75}{l}{#4}%
\cell{2.2}{1.25}{l}{#5}%
\qbezier(0,3)(2,7)(4,3)%
\qbezier(0,2)(2,-2)(4,2)%
\put(0,2.5){\vector(1,0){4}}%
\put(2,4.5){\vector(0,-1){1.5}}%
\put(2,2){\vector(0,-1){1.5}}%
\put(4,3){\vector(1,-3){0}}%
\put(4,2){\vector(1,3){0}}%
\end{picture}}

\mcm{\cthree}{5}{%
\cinitdims{4.2}{5.4}%
\abovepic{#1}%
\belowpic{#3}%
\present{\precthree{#1}{#2}{#3}{#4}{#5}}}

\newcommand{\prectwoop}[3]%
{\begin{picture}(4.2,3.4)(-0.1,-0.2)%
\cell{2}{3.2}{b}{#1}%
\cell{2}{-0.2}{t}{#2}%
\cell{2.2}{1.5}{l}{#3}%
\qbezier(0,2)(2,4)(4,2)%
\qbezier(0,1)(2,-1)(4,1)%
\put(0,2){\vector(-1,-1){0}}%
\put(0,1){\vector(-1,1){0}}%
\put(2,2.5){\vector(0,-1){2}}%
\end{picture}}

\mcm{\ctwoop}{3}{%
\cinitdims{4.2}{3.4}%
\abovepic{#1}%
\belowpic{#2}%
\present{\prectwoop{#1}{#2}{#3}}}

\newcommand{\prectwopar}[4]{%
\begin{picture}(4.2,3.4)(-0.1,-0.2)%
\cell{2}{3.2}{b}{#1}%
\cell{2}{-0.2}{t}{#2}%
\cell{1.6}{1.5}{r}{#3}%
\cell{2.4}{1.5}{l}{#4}%
\qbezier(0,2)(2,4)(4,2)%
\qbezier(0,1)(2,-1)(4,1)%
\put(4,2){\vector(1,-1){0}}%
\put(4,1){\vector(1,1){0}}%
\put(1.8,2.5){\vector(0,-1){2}}%
\put(2.2,2.5){\vector(0,-1){2}}%
\end{picture}}

\mcm{\ctwopar}{4}{%
\cinitdims{4.2}{3.4}%
\abovepic{#1}%
\belowpic{#2}%
\present{\prectwopar{#1}{#2}{#3}{#4}}}

\newcommand{\precthreein}[5]{%
\begin{picture}(4.2,5.4)(-0.1,-0.2)%
\cell{2}{5.2}{b}{#1}%
\cell{1}{2.7}{b}{#2}%
\cell{2}{-.2}{t}{#3}%
\cell{2.2}{3.75}{l}{#4}%
\cell{2.2}{1.25}{l}{#5}%
\qbezier(0,3)(2,7)(4,3)%
\qbezier(0,2)(2,-2)(4,2)%
\put(0,2.5){\vector(1,0){4}}%
\put(2,4.5){\vector(0,-1){1.5}}%
\put(2,0.5){\vector(0,1){1.5}}%
\put(4,3){\vector(1,-3){0}}%
\put(4,2){\vector(1,3){0}}%
\end{picture}}

\mcm{\cthreein}{5}{%
\cinitdims{4.2}{5.4}%
\abovepic{#1}%
\belowpic{#3}%
\present{\precthreein{#1}{#2}{#3}{#4}{#5}}}

\newcommand{\precthreecell}[5]{%
\begin{picture}(8.2,5)(-4.1,-2.5)%
\cell{0}{2.5}{b}{#1}%
\cell{0}{-2.5}{t}{#2}%
\cell{-1.7}{0}{r}{#3}%
\cell{1.7}{0}{l}{#4}%
\cell{0}{0.2}{b}{#5}%
\qbezier(-4,0.5)(0,4)(4,0.5)%
\qbezier(-4,-0.5)(0,-4)(4,-0.5)%
\qbezier(-0.5,2)(-2.5,0)(-0.5,-2)%
\qbezier(0.5,2)(2.5,0)(0.5,-2)%
\put(-1,0){\vector(1,0){2}}%
\put(4,0.5){\vector(1,-1){0}}%
\put(4,-0.5){\vector(1,1){0}}%
\put(-0.5,-2){\vector(1,-1){0}}%
\put(0.5,-2){\vector(-1,-1){0}}%
\end{picture}}

\mcm{\cthreecell}{5}{%
\cinitdims{8.2}{5}%
\abovepic{#1}%
\belowpic{#2}%
\present{\precthreecell{#1}{#2}{#3}{#4}{#5}}}

\newcommand{\precthreecellpar}[6]{%
\begin{picture}(8.2,5)(-4.1,-2.5)%
\cell{0}{2.5}{b}{#1}%
\cell{0}{-2.5}{t}{#2}%
\cell{-1.7}{0}{r}{#3}%
\cell{1.7}{0}{l}{#4}%
\cell{0}{0.4}{b}{#5}%
\cell{0}{-0.4}{t}{#6}%
\qbezier(-4,0.5)(0,4)(4,0.5)%
\qbezier(-4,-0.5)(0,-4)(4,-0.5)%
\qbezier(-0.5,2)(-2.5,0)(-0.5,-2)%
\qbezier(0.5,2)(2.5,0)(0.5,-2)%
\put(-1,0.2){\vector(1,0){2}}%
\put(-1,-0.2){\vector(1,0){2}}%
\put(4,0.5){\vector(1,-1){0}}%
\put(4,-0.5){\vector(1,1){0}}%
\put(-0.5,-2){\vector(1,-1){0}}%
\put(0.5,-2){\vector(-1,-1){0}}%
\end{picture}}

\mcm{\cthreecellpar}{6}{%
\cinitdims{8.2}{5}%
\abovepic{#1}%
\belowpic{#2}%
\present{\precthreecellpar{#1}{#2}{#3}{#4}{#5}{#6}}}

\newcommand{\prectwov}[5]{%
\begin{picture}(3.4,4.2)(0.8,0.9)%
\cell{2.5}{5.1}{b}{#1}%
\cell{2.5}{0.9}{t}{#2}%
\cell{0.8}{3}{r}{#3}%
\cell{4.2}{3}{l}{#4}%
\cell{2.5}{3.2}{b}{#5}%
\qbezier(2,5)(0,3)(2,1)%
\qbezier(3,5)(5,3)(3,1)%
\put(2,1){\vector(1,-1){0}}%
\put(3,1){\vector(-1,-1){0}}%
\put(1.5,3){\vector(1,0){2}}%
\end{picture}}

\mcm{\ctwov}{5}{%
\cinitdims{3.4}{4.2}%
\abovepic{#1}%
\belowpic{#2}%
\sidespic{#3}%
\sidespic{#4}%
\present{\prectwov{#1}{#2}{#3}{#4}{#5}}}

\newcommand{\precthreecellv}[7]{%
\begin{picture}(5,8.2)(0.5,-1.6)%
\cell{3}{6.6}{b}{#1}%
\cell{3}{-1.6}{t}{#2}%
\cell{0.5}{2.5}{r}{#3}%
\cell{5.5}{2.5}{l}{#4}%
\cell{3}{4.2}{b}{#5}%
\cell{3}{0.8}{t}{#6}%
\cell{3.2}{2.5}{l}{#7}%
\qbezier(3.5,6.5)(7,2.5)(3.5,-1.5)%
\qbezier(2.5,6.5)(-1,2.5)(2.5,-1.5)%
\put(2.5,-1.5){\vector(1,-1){0}}%
\put(3.5,-1.5){\vector(-1,-1){0}}%
\qbezier(1,3)(3,5)(5,3)%
\qbezier(1,2)(3,0)(5,2)%
\put(5,3){\vector(1,-1){0}}%
\put(5,2){\vector(1,1){0}}%
\put(3,3.5){\vector(0,-1){2}}%
\end{picture}}

\mcm{\cthreecellv}{7}{%
\cinitdims{5}{8.2}%
\abovepic{#1}%
\belowpic{#2}%
\sidespic{#3}%
\sidespic{#4}%
\present{\precthreecellv{#1}{#2}{#3}{#4}{#5}{#6}{#7}}}

\newcommand{\pretopez}[2]{%
\begin{picture}(2.6,2.3)(-1.3,-2.2)%
\cell{0}{-2.2}{t}{#1}%
\cell{0}{-1.2}{c}{#2}%
\qbezier(0,0)(-2,-2)(0,-2)%
\qbezier(0,0)(2,-2)(0,-2)%
\put(0,0){\vector(-1,1){0}}%
\end{picture}}

\mcm{\topez}{2}{%
\ginitdims{2.6}{2.3}%
\belowpic{#1}%
\present{\pretopez{#1}{#2}}}

\newcommand{\pretopea}[3]{%
\begin{picture}(4,1.9)(-2,-0,2)%
\cell{0}{1.7}{b}{#1}%
\cell{0}{-0.2}{t}{#2}%
\cell{0}{0.7}{c}{#3}%
\qbezier(-2,0)(0,3)(2,0)%
\put(-2,0){\vector(1,0){4}}%
\put(2,0){\vector(2,-3){0}}%
\end{picture}}

\mcm{\topea}{3}{%
\ginitdims{4}{1.9}%
\abovepic{#1}%
\belowpic{#2}%
\present{\pretopea{#1}{#2}{#3}}}

\newcommand{\pretopeb}[4]{%
\begin{picture}(4,2.2)(-2,-0.2)%
\cell{-1.1}{1}{br}{#1}%
\cell{1.1}{1}{bl}{#2}%
\cell{0}{-0.2}{t}{#3}%
\cell{0}{0.8}{c}{#4}%
\put(-2,0){\vector(1,1){2}}%
\put(0,2){\vector(1,-1){2}}%
\put(-2,0){\vector(1,0){4}}%
\end{picture}}

\mcm{\topeb}{4}{%
\ginitdims{4}{2.2}%
\belowpic{#3}%
\present{\pretopeb{#1}{#2}{#3}{#4}}}

\newcommand{\pretopec}[5]{%
\begin{picture}(4,2.2)(-2,-0.2)%
\cell{-1.8}{1}{br}{#1}%
\cell{0}{2.2}{b}{#2}%
\cell{1.8}{1}{bl}{#3}%
\cell{0}{-0.2}{t}{#4}%
\cell{0}{0.8}{c}{#5}%
\put(-2,0){\vector(1,2){1}}%
\put(-1,2){\vector(1,0){2}}%
\put(1,2){\vector(1,-2){1}}%
\put(-2,0){\vector(1,0){4}}%
\end{picture}}

\mcm{\topec}{5}{%
\ginitdims{4}{2.2}%
\sidespic{#1}%
\abovepic{#2}%
\sidespic{#3}%
\belowpic{#4}%
\present{\pretopec{#1}{#2}{#3}{#4}{#5}}}

\newcommand{\pretoped}[6]{%
\begin{picture}(4,2.5)(-2,-0.2)%
\cell{-2}{0.6}{br}{#1}%
\cell{-0.7}{2.2}{br}{#2}%
\cell{0.7}{2.2}{bl}{#3}%
\cell{2}{0.6}{bl}{#4}%
\cell{0}{-0.2}{t}{#5}%
\cell{0}{0.8}{c}{#6}%
\put(-2,0){\vector(1,3){0.5}}%
\put(-1.5,1.5){\vector(3,2){1.5}}%
\put(0,2.5){\vector(3,-2){1.5}}%
\put(1.5,1.5){\vector(1,-3){0.5}}%
\put(-2,0){\vector(1,0){4}}%
\end{picture}}

\mcm{\toped}{6}{%
\ginitdims{4}{2.5}%
\sidespic{#1}%
\abovepic{#2}%
\abovepic{#3}%
\sidespic{#4}%
\belowpic{#5}%
\present{\pretoped{#1}{#2}{#3}{#4}{#5}{#6}}}

\newcommand{\pretopeq}[5]{%
\begin{picture}(4,2.5)(-2,-0.2)%
\cell{-2}{0.6}{br}{#1}%
\cell{-1}{2.2}{br}{#2}%
\cell{2}{0.6}{bl}{#3}%
\cell{0}{-0.2}{t}{#4}%
\cell{0}{0.8}{c}{#5}%
\put(-2,0){\vector(1,3){0.5}}%
\put(-1.5,1.5){\vector(1,1){1}}%
\cell{0.9}{2.3}{c}{\ddots}
\put(1.5,1.5){\vector(1,-3){0.5}}%
\put(-2,0){\vector(1,0){4}}%
\end{picture}}

\mcm{\topeq}{5}{%
\ginitdims{4}{2.5}%
\sidespic{#1}%
\abovepic{#2}%
\sidespic{#3}%
\belowpic{#4}%
\present{\pretopeq{#1}{#2}{#3}{#4}{#5}}}

\newcommand{\pretopebase}[1]{%
\begin{picture}(4,0.4)(0,-0.2)%
\cell{2}{0.2}{b}{#1}%
\put(0,0){\vector(1,0){4}}%
\end{picture}}

\mcm{\topebase}{1}{%
\ginitdims{4}{0.4}%
\abovepic{#1}%
\present{\pretopebase{#1}}}

\newcommand{\pretopezs}[2]{%
\begin{picture}(2.6,2.3)(-1.3,-2.2)%
\cell{0}{-2.2}{t}{#1}%
\cell{0}{-1.2}{c}{#2}%
\qbezier(0,0)(-2,-2)(0,-2)%
\qbezier(0,0)(2,-2)(0,-2)%
\end{picture}}

\mcm{\topezs}{2}{%
\ginitdims{2.6}{2.3}%
\belowpic{#1}%
\present{\pretopezs{#1}{#2}}}

\newcommand{\pretopeas}[3]{%
\begin{picture}(4,1.9)(-2,-0,2)%
\cell{0}{1.7}{b}{#1}%
\cell{0}{-0.2}{t}{#2}%
\cell{0}{0.7}{c}{#3}%
\qbezier(-2,0)(0,3)(2,0)%
\put(-2,0){\line(1,0){4}}%
\end{picture}}

\mcm{\topeas}{3}{%
\ginitdims{4}{1.9}%
\abovepic{#1}%
\belowpic{#2}%
\present{\pretopeas{#1}{#2}{#3}}}

\newcommand{\pretopebs}[4]{%
\begin{picture}(4,2.2)(-2,-0.2)%
\cell{-1.1}{1}{br}{#1}%
\cell{1.1}{1}{bl}{#2}%
\cell{0}{-0.2}{t}{#3}%
\cell{0}{0.8}{c}{#4}%
\put(-2,0){\line(1,1){2}}%
\put(0,2){\line(1,-1){2}}%
\put(-2,0){\line(1,0){4}}%
\end{picture}}

\mcm{\topebs}{4}{%
\ginitdims{4}{2.2}%
\belowpic{#3}%
\present{\pretopebs{#1}{#2}{#3}{#4}}}

\newcommand{\pretopecs}[5]{%
\begin{picture}(4,2.2)(-2,-0.2)%
\cell{-1.8}{1}{br}{#1}%
\cell{0}{2.2}{b}{#2}%
\cell{1.8}{1}{bl}{#3}%
\cell{0}{-0.2}{t}{#4}%
\cell{0}{0.8}{c}{#5}%
\put(-2,0){\line(1,2){1}}%
\put(-1,2){\line(1,0){2}}%
\put(1,2){\line(1,-2){1}}%
\put(-2,0){\line(1,0){4}}%
\end{picture}}

\mcm{\topecs}{5}{%
\ginitdims{4}{2.2}%
\sidespic{#1}%
\abovepic{#2}%
\sidespic{#3}%
\belowpic{#4}%
\present{\pretopecs{#1}{#2}{#3}{#4}{#5}}}

\newcommand{\pretopeds}[6]{%
\begin{picture}(4,2.5)(-2,-0.2)%
\cell{-2}{0.6}{br}{#1}%
\cell{-0.7}{2.2}{br}{#2}%
\cell{0.7}{2.2}{bl}{#3}%
\cell{2}{0.6}{bl}{#4}%
\cell{0}{-0.2}{t}{#5}%
\cell{0}{0.8}{c}{#6}%
\put(-2,0){\line(1,3){0.5}}%
\put(-1.5,1.5){\line(3,2){1.5}}%
\put(0,2.5){\line(3,-2){1.5}}%
\put(1.5,1.5){\line(1,-3){0.5}}%
\put(-2,0){\line(1,0){4}}%
\end{picture}}

\mcm{\topeds}{6}{%
\ginitdims{4}{2.5}%
\sidespic{#1}%
\abovepic{#2}%
\abovepic{#3}%
\sidespic{#4}%
\belowpic{#5}%
\present{\pretopeds{#1}{#2}{#3}{#4}{#5}{#6}}}

\newcommand{\pretopeqs}[5]{%
\begin{picture}(4,2.5)(-2,-0.2)%
\cell{-2}{0.6}{br}{#1}%
\cell{-1}{2.2}{br}{#2}%
\cell{2}{0.6}{bl}{#3}%
\cell{0}{-0.2}{t}{#4}%
\cell{0}{0.8}{c}{#5}%
\put(-2,0){\line(1,3){0.5}}%
\put(-1.5,1.5){\line(1,1){1}}%
\cell{0.9}{2.3}{c}{\ddots}
\put(1.5,1.5){\line(1,-3){0.5}}%
\put(-2,0){\line(1,0){4}}%
\end{picture}}

\mcm{\topeqs}{5}{%
\ginitdims{4}{2.5}%
\sidespic{#1}%
\abovepic{#2}%
\sidespic{#3}%
\belowpic{#4}%
\present{\pretopeqs{#1}{#2}{#3}{#4}{#5}}}

\newcommand{\pretopebases}[1]{%
\begin{picture}(4,0.4)(0,-0.2)%
\cell{2}{0.2}{b}{#1}%
\put(0,0){\line(1,0){4}}%
\end{picture}}

\mcm{\topebases}{1}{%
\ginitdims{4}{0.4}%
\abovepic{#1}%
\present{\pretopebases{#1}}}

\newcommand{\pregdots}[6]{%
\begin{picture}(5,8.4)(0,-2.7)%
\cell{2.5}{5.7}{b}{#1}%
\cell{1.5}{2.8}{b}{#2}%
\cell{1.5}{0.2}{t}{#3}%
\cell{2.5}{-2.7}{t}{#4}%
\cell{2.7}{4.25}{l}{#5}%
\cell{2.7}{-1.25}{l}{#6}%
\qbezier(0,1.5)(2.5,9.5)(5,1.5)%
\qbezier(0,1.5)(2.5,4)(5,1.5)%
\qbezier(0,1.5)(2.5,-1)(5,1.5)%
\qbezier(0,1.5)(2.5,-6.5)(5,1.5)%
\put(2.5,5.25){\vector(0,-1){2}}%
\put(2.5,-0.25){\vector(0,-1){2}}%
\cell{2.5}{1.7}{c}{\vdots}%
\put(5,1.5){\vector(1,-4){0}}%
\put(5,1.5){\vector(4,-3){0}}%
\put(5,1.5){\vector(4,3){0}}%
\put(5,1.5){\vector(1,4){0}}%
\end{picture}}

\mcm{\gdots}{6}{%
\ginitdims{5}{8.4}%
\abovepic{#1}%
\belowpic{#4}%
\present{\pregdots{#1}{#2}{#3}{#4}{#5}{#6}}}

\newlength{\volt}
\makeatletter

\def\diagram{\m@th\leftwidth=\z@ \rightwidth=\z@ \topheight=\z@
\botheight=\z@ \setbox\@picbox\hbox\bgroup}

\def\enddiagram{\egroup\wd\@picbox\rightwidth\unitlength
\ht\@picbox\topheight\unitlength \dp\@picbox\botheight\unitlength
\hskip\leftwidth\unitlength\box\@picbox}

\def\bfig{\begin{diagram}}
\def\efig{\end{diagram}}
\newcount\wideness \newcount\leftwidth \newcount\rightwidth
\newcount\highness \newcount\topheight \newcount\botheight

\def\ratchet#1#2{\ifnum#1<#2 \global #1=#2 \fi}

\def\putbox(#1,#2)#3{%
\horsize{\wideness}{#3} \divide\wideness by 2 {\advance\wideness
by #1 \ratchet{\rightwidth}{\wideness}} {\advance\wideness by -#1
\ratchet{\leftwidth}{\wideness}} \vertsize{\highness}{#3}
\divide\highness by 2 {\advance\highness by #2
\ratchet{\topheight}{\highness}} {\advance\highness by -#2
\ratchet{\botheight}{\highness}} \put(#1,#2){\makebox(0,0){$#3$}}}

\def\putlbox(#1,#2)#3{%
\horsize{\wideness}{#3} {\advance\wideness by #1
\ratchet{\rightwidth}{\wideness}} {\ratchet{\leftwidth}{-#1}}
\vertsize{\highness}{#3} \divide\highness by 2 {\advance\highness
by #2 \ratchet{\topheight}{\highness}} {\advance\highness by -#2
\ratchet{\botheight}{\highness}}
\put(#1,#2){\makebox(0,0)[l]{$#3$}}}

\def\putrbox(#1,#2)#3{%
\horsize{\wideness}{#3} {\ratchet{\rightwidth}{#1}}
{\advance\wideness by -#1 \ratchet{\leftwidth}{\wideness}}
\vertsize{\highness}{#3} \divide\highness by 2 {\advance\highness
by #2 \ratchet{\topheight}{\highness}} {\advance\highness by -#2
\ratchet{\botheight}{\highness}}
\put(#1,#2){\makebox(0,0)[r]{$#3$}}}

\def\adjust[#1]{} 

\newcount \coefa
\newcount \coefb
\newcount \coefc
\newcount\tempcounta
\newcount\tempcountb
\newcount\tempcountc
\newcount\tempcountd
\newcount\xext
\newcount\yext
\newcount\xoff
\newcount\yoff
\newcount\gap%
\newcount\arrowtypea
\newcount\arrowtypeb
\newcount\arrowtypec
\newcount\arrowtyped
\newcount\arrowtypee
\newcount\height
\newcount\width
\newcount\xpos
\newcount\ypos
\newcount\run
\newcount\rise
\newcount\arrowlength
\newcount\halflength
\newcount\arrowtype
\newdimen\tempdimen
\newdimen\xlen
\newdimen\ylen
\newsavebox{\tempboxa}%
\newsavebox{\tempboxb}%
\newsavebox{\tempboxc}%

\newdimen\w@dth

\def\setw@dth#1#2{\setbox\z@\hbox{\m@th$#1$}\w@dth=\wd\z@
\setbox\@ne\hbox{\m@th$#2$}\ifnum\w@dth<\wd\@ne \w@dth=\wd\@ne \fi
\advance\w@dth by 1.2em}

\def\t@^#1_#2{\allowbreak\def\n@one{#1}\def\n@two{#2}\mathrel
{\setw@dth{#1}{#2} \mathop{\hbox to
\w@dth{\rightarrowfill}}\limits \ifx\n@one\empty\else
^{\box\z@}\fi \ifx\n@two\empty\else _{\box\@ne}\fi}}
\def\t@@^#1{\@ifnextchar_{\t@^{#1}}{\t@^{#1}_{}}}
\def\to{\@ifnextchar^{\t@@}{\t@@^{}}}

\def\t@left^#1_#2{\def\n@one{#1}\def\n@two{#2}\mathrel{\setw@dth{#1}{#2}
\mathop{\hbox to \w@dth{\leftarrowfill}}\limits
\ifx\n@one\empty\else ^{\box\z@}\fi \ifx\n@two\empty\else
_{\box\@ne}\fi}}
\def\t@@left^#1{\@ifnextchar_{\t@left^{#1}}{\t@left^{#1}_{}}}
\def\toleft{\@ifnextchar^{\t@@left}{\t@@left^{}}}

\def\two@^#1_#2{\allowbreak
\def\n@one{#1}\def\n@two{#2}\mathrel{\setw@dth{#1}{#2}
\mathop{\vcenter{\lineskip\z@\baselineskip\z@
                 \hbox to \w@dth{\rightarrowfill}%
                 \hbox to \w@dth{\rightarrowfill}}%
       }\limits
\ifx\n@one\empty\else ^{\box\z@}\fi \ifx\n@two\empty\else
_{\box\@ne}\fi}}
\def\tw@@^#1{\@ifnextchar _{\two@^{#1}}{\two@^{#1}_{}}}
\def\two{\@ifnextchar ^{\tw@@}{\tw@@^{}}}

\def\tofr@^#1_#2{\def\n@one{#1}\def\n@two{#2}\mathrel{\setw@dth{#1}{#2}
\mathop{\vcenter{\hbox to \w@dth{\rightarrowfill}\kern-1.7ex
                 \hbox to \w@dth{\leftarrowfill}}%
       }\limits
\ifx\n@one\empty\else ^{\box\z@}\fi \ifx\n@two\empty\else
_{\box\@ne}\fi}}
\def\t@fr@^#1{\@ifnextchar_ {\tofr@^{#1}}{\tofr@^{#1}_{}}}
\def\tofro{\@ifnextchar^ {\t@fr@}{\t@fr@^{}}}

\def\mon{\mathop{\m@th\hbox to
      14.6\P@{\lasyb\char'51\hskip-2.1\P@$\arrext$\hss
$\mathord\rightarrow$}}\limits} 
\def\leftmono{\mathrel{\m@th\hbox to
14.6\P@{$\mathord\leftarrow$\hss$\arrext$\hskip-2.1\P@\lasyb\char'50%
}}\limits} 
\mathchardef\arrext="0200       

\def\settypes(#1,#2,#3){\arrowtypea#1 \arrowtypeb#2 \arrowtypec#3}
\def\settoheight#1#2{\setbox\@tempboxa\hbox{#2}#1\ht\@tempboxa\relax}%
\def\settodepth#1#2{\setbox\@tempboxa\hbox{#2}#1\dp\@tempboxa\relax}%
\def\settokens`#1`#2`#3`#4`{%
     \def\tokena{#1}\def\tokenb{#2}\def\tokenc{#3}\def\tokend{#4}}
\def\setsqparms[#1`#2`#3`#4;#5`#6]{%
\arrowtypea #1 \arrowtypeb #2 \arrowtypec #3 \arrowtyped #4
\width #5 \height #6 }
\def\setpos(#1,#2){\xpos=#1 \ypos#2}

\def\settriparms[#1`#2`#3;#4]{\settripairparms[#1`#2`#3`1`1;#4]}%

\def\settripairparms[#1`#2`#3`#4`#5;#6]{%
\arrowtypea #1 \arrowtypeb #2 \arrowtypec #3 \arrowtyped #4
\arrowtypee #5 \width #6 \height #6 }

\def\resetparms{\settripairparms[1`1`1`1`1;500]\width 500}

\resetparms

\def\mvector(#1,#2)#3{
\put(0,0){\vector(#1,#2){#3}}%
\put(0,0){\vector(#1,#2){26}}%
}
\def\evector(#1,#2)#3{{
\arrowlength #3
\put(0,0){\vector(#1,#2){\arrowlength}}%
\advance \arrowlength by-30
\put(0,0){\vector(#1,#2){\arrowlength}}%
}}

\def\horsize#1#2{%
\settowidth{\tempdimen}{$#2$}%
#1=\tempdimen \divide #1 by\unitlength }

\def\vertsize#1#2{%
\settoheight{\tempdimen}{$#2$}%
#1=\tempdimen
\settodepth{\tempdimen}{$#2$}%
\advance #1 by\tempdimen \divide #1 by\unitlength }

\def\putvector(#1,#2)(#3,#4)#5#6{{%
\ifnum3<\arrowtype \putdashvector(#1,#2)(#3,#4)#5\arrowtype \else
\ifnum\arrowtype<-3 \putdashvector(#1,#2)(#3,#4)#5\arrowtype \else
\xpos=#1 \ypos=#2 \run=#3 \rise=#4 \arrowlength=#5 \ifnum
\arrowtype<0
    \ifnum \run=0
        \advance \ypos by-\arrowlength
    \else
        \tempcounta \arrowlength
        \multiply \tempcounta by\rise
        \divide \tempcounta by\run
        \ifnum\run>0
            \advance \xpos by\arrowlength
            \advance \ypos by\tempcounta
        \else
            \advance \xpos by-\arrowlength
            \advance \ypos by-\tempcounta
        \fi
    \fi
    \multiply \arrowtype by-1
    \multiply \rise by-1
    \multiply \run by-1
\fi \ifcase \arrowtype
\or \put(\xpos,\ypos){\vector(\run,\rise){\arrowlength}}%
\or \put(\xpos,\ypos){\mvector(\run,\rise)\arrowlength}%
\or \put(\xpos,\ypos){\evector(\run,\rise){\arrowlength}}%
\fi\fi\fi }}

\def\putsplitvector(#1,#2)#3#4{
\xpos #1 \ypos #2 \arrowtype #4 \halflength #3 \arrowlength #3
\gap 140 \advance \halflength by-\gap \divide \halflength by2
\ifnum\arrowtype>0
   \ifcase \arrowtype
   \or \put(\xpos,\ypos){\line(0,-1){\halflength}}%
       \advance\ypos by-\halflength
       \advance\ypos by-\gap
       \put(\xpos,\ypos){\vector(0,-1){\halflength}}%
   \or \put(\xpos,\ypos){\line(0,-1)\halflength}%
       \put(\xpos,\ypos){\vector(0,-1)3}%
       \advance\ypos by-\halflength
       \advance\ypos by-\gap
       \put(\xpos,\ypos){\vector(0,-1){\halflength}}%
   \or \put(\xpos,\ypos){\line(0,-1)\halflength}%
       \advance\ypos by-\halflength
       \advance\ypos by-\gap
       \put(\xpos,\ypos){\evector(0,-1){\halflength}}%
   \fi
\else \arrowtype=-\arrowtype
   \ifcase\arrowtype
   \or \advance \ypos by-\arrowlength
       \put(\xpos,\ypos){\line(0,1){\halflength}}%
       \advance\ypos by\halflength
       \advance\ypos by\gap
       \put(\xpos,\ypos){\vector(0,1){\halflength}}%
   \or \advance \ypos by-\arrowlength
       \put(\xpos,\ypos){\line(0,1)\halflength}%
       \put(\xpos,\ypos){\vector(0,1)3}%
       \advance\ypos by\halflength
       \advance\ypos by\gap
       \put(\xpos,\ypos){\vector(0,1){\halflength}}%
   \or \advance \ypos by-\arrowlength
       \put(\xpos,\ypos){\line(0,1)\halflength}%
       \advance\ypos by\halflength
       \advance\ypos by\gap
       \put(\xpos,\ypos){\evector(0,1){\halflength}}%
   \fi
\fi }

\def\putmorphism(#1)(#2,#3)[#4`#5`#6]#7#8#9{{%
\run #2 \rise #3 \ifnum\rise=0
  \puthmorphism(#1)[#4`#5`#6]{#7}{#8}#9%
\else\ifnum\run=0
  \putvmorphism(#1)[#4`#5`#6]{#7}{#8}#9%
\else
\setpos(#1)%
\arrowlength #7 \arrowtype #8 \ifnum\run=0 \else\ifnum\rise=0
\else \ifnum\run>0
    \coefa=1
\else
   \coefa=-1
\fi \ifnum\arrowtype>0
   \coefb=0
   \coefc=-1
\else
   \coefb=\coefa
   \coefc=1
   \arrowtype=-\arrowtype
\fi \width=2 \multiply \width by\run \divide \width by\rise
\ifnum \width<0  \width=-\width\fi \advance\width by60 \if l#9
\width=-\width\fi
\putbox(\xpos,\ypos){#4}
{\multiply \coefa by\arrowlength
\advance\xpos by\coefa \multiply \coefa by\rise \divide \coefa
by\run \advance \ypos by\coefa
\putbox(\xpos,\ypos){#5} }%
{\multiply \coefa by\arrowlength
\divide \coefa by2 \advance \xpos by\coefa \advance \xpos by\width
\multiply \coefa by\rise \divide \coefa by\run \advance \ypos
by\coefa
\if l#9%
   \putrbox(\xpos,\ypos){#6}%
\else\if r#9%
   \putlbox(\xpos,\ypos){#6}%
\fi\fi }%
{\multiply \rise by-\coefc
\multiply \run by-\coefc \multiply \coefb by\arrowlength \advance
\xpos by\coefb \multiply \coefb by\rise \divide \coefb by\run
\advance \ypos by\coefb \multiply \coefc by70 \advance \ypos
by\coefc \multiply \coefc by\run \divide \coefc by\rise \advance
\xpos by\coefc \multiply \coefa by140 \multiply \coefa by\run
\divide \coefa by\rise \advance \arrowlength by\coefa
\ifcase\arrowtype
\or \put(\xpos,\ypos){\vector(\run,\rise){\arrowlength}}%
\or \put(\xpos,\ypos){\mvector(\run,\rise){\arrowlength}}%
\or \put(\xpos,\ypos){\evector(\run,\rise){\arrowlength}}%
\fi}\fi\fi\fi\fi}}

\newcount\numbdashes \newcount\lengthdash \newcount\increment

\def\howmanydashes{
\numbdashes=\arrowlength \lengthdash=40 \divide\numbdashes by
\lengthdash \lengthdash=\arrowlength \divide\lengthdash by
\numbdashes
\increment=\lengthdash \multiply\lengthdash by 3
\divide\lengthdash by 5 }

\def\putdashvector(#1)(#2,#3)#4#5{%
\ifnum#3=0 \putdashhvector(#1){#4}#5 \else \ifnum#2=0
\putdashvvector(#1){#4}#5\fi\fi}

\def\putdashhvector(#1,#2)#3#4{{%
\arrowlength=#3 \howmanydashes
\multiput(#1,#2)(\increment,0){\numbdashes}%
{\vrule height .4pt width \lengthdash\unitlength} \arrowtype=#4
\xpos=#1 \ifnum\arrowtype<0 \advance\arrowtype by 7 \fi
\ifcase\arrowtype \or \advance\xpos by 10
    \put(\xpos,#2){\vector(-1,0){\lengthdash}}
    \advance\xpos by 40
    \put(\xpos,#2){\vector(-1,0){\lengthdash}}
\or \advance \xpos by 10
    \put(\xpos,#2){\vector(-1,0){\lengthdash}}
    \advance\xpos by  \arrowlength
    \advance\xpos by  -50
    \put(\xpos,#2){\vector(-1,0){\lengthdash}}
\or \advance\xpos by 10
    \put(\xpos,#2){\vector(-1,0){\lengthdash}}
\or \advance\xpos by \arrowlength
    \advance\xpos by -\lengthdash
    \put(\xpos,#2){\vector(1,0){\lengthdash}}
\or {\advance\xpos by 10
    \put(\xpos,#2){\vector(1,0){\lengthdash}}}
    \advance\xpos by \arrowlength
    \advance\xpos by -\lengthdash
    \put(\xpos,#2){\vector(1,0){\lengthdash}}
\or \advance\xpos by \arrowlength
    \advance\xpos by -\lengthdash
    \put(\xpos,#2){\vector(1,0){\lengthdash}}
    \advance\xpos by -40
    \put(\xpos,#2){\vector(1,0){\lengthdash}}
   \fi
}}

\def\putdashvvector(#1,#2)#3#4{{%
\arrowlength=#3 \howmanydashes \ypos=#2 \advance\ypos by
-\arrowlength
\multiput(#1,#2)(0,\increment){\numbdashes}%
    {\vrule width .4pt height \lengthdash\unitlength}
\arrowtype=#4 \ypos=#2 \ifnum\arrowtype<0 \advance\arrowtype by 7
\fi \ifcase\arrowtype \or \advance\ypos by \arrowlength
\advance\ypos by -40
    \put(#1,\ypos){\vector(0,1){\lengthdash}}
    \advance\ypos by -40
    \put(#1,\ypos){\vector(0,1){\lengthdash}}
\or \advance\ypos by 10
    \put(#1,\ypos){\vector(0,1){\lengthdash}}
    \advance\ypos by \arrowlength \advance\ypos by -40
    \put(#1,\ypos){\vector(0,1){\lengthdash}}
\or \advance\ypos by \arrowlength \advance\ypos by -40
    \put(#1,\ypos){\vector(0,1){\lengthdash}}
\or \advance\ypos by 10
    \put(#1,\ypos){\vector(0,-1){\lengthdash}}
\or \advance\ypos by 10
    \put(#1,\ypos){\vector(0,-1){\lengthdash}}
    \advance\ypos by \arrowlength \advance\ypos by -40
    \put(#1,\ypos){\vector(0,-1){\lengthdash}}
\or \advance\ypos by 10
    \put(#1,\ypos){\vector(0,-1){\lengthdash}}
    \advance\ypos by 40
    \put(#1,\ypos){\vector(0,-1){\lengthdash}}
\fi }}

\def\puthmorphism(#1,#2)[#3`#4`#5]#6#7#8{{%
\xpos #1 \ypos #2 \width #6 \arrowlength #6 \arrowtype=#7
\putbox(\xpos,\ypos){#3\vphantom{#4}}%
{\advance \xpos by\arrowlength
\putbox(\xpos,\ypos){\vphantom{#3}#4}}%
\horsize{\tempcounta}{#3}%
\horsize{\tempcountb}{#4}%
\divide \tempcounta by2 \divide \tempcountb by2 \advance
\tempcounta by30 \advance \tempcountb by30 \advance \xpos
by\tempcounta \advance \arrowlength by-\tempcounta \advance
\arrowlength by-\tempcountb
\putvector(\xpos,\ypos)(1,0)\arrowlength\arrowtype \divide
\arrowlength by2 \advance \xpos by\arrowlength
\vertsize{\tempcounta}{#5}%
\divide\tempcounta by2 \advance \tempcounta by20
\if a#8 %
   \advance \ypos by\tempcounta
   \putbox(\xpos,\ypos){#5}%
\else
   \advance \ypos by-\tempcounta
   \putbox(\xpos,\ypos){#5}%
\fi}}

\def\putvmorphism(#1,#2)[#3`#4`#5]#6#7#8{{%
\xpos #1 \ypos #2 \arrowlength #6 \arrowtype #7
\settowidth{\xlen}{$#5$}%
\putbox(\xpos,\ypos){#3}%
{\advance \ypos by-\arrowlength
\putbox(\xpos,\ypos){#4}}%
{\advance\arrowlength by-140 \advance \ypos by-70 \ifdim\xlen>0pt
   \if m#8%
      \putsplitvector(\xpos,\ypos)\arrowlength\arrowtype
   \else
   \putvector(\xpos,\ypos)(0,-1)\arrowlength\arrowtype
   \fi
\else
   \putvector(\xpos,\ypos)(0,-1)\arrowlength\arrowtype
\fi}%
\ifdim\xlen>0pt
   \divide \arrowlength by2
   \advance\ypos by-\arrowlength
   \if l#8%
      \advance \xpos by-40
      \putrbox(\xpos,\ypos){#5}%
   \else\if r#8%
      \advance \xpos by40
      \putlbox(\xpos,\ypos){#5}%
   \else
      \putbox(\xpos,\ypos){#5}%
   \fi\fi
\fi }}

\def\putsquarep<#1>(#2)[#3;#4`#5`#6`#7]{{%
\setsqparms[#1]%
\setpos(#2)%
\settokens`#3`%
\puthmorphism(\xpos,\ypos)[\tokenc`\tokend`{#7}]{\width}{\arrowtyped}b%
\advance\ypos by \height
\puthmorphism(\xpos,\ypos)[\tokena`\tokenb`{#4}]{\width}{\arrowtypea}a%
\putvmorphism(\xpos,\ypos)[``{#5}]{\height}{\arrowtypeb}l%
\advance\xpos by \width
\putvmorphism(\xpos,\ypos)[``{#6}]{\height}{\arrowtypec}r%
}}

\def\putsquare{\@ifnextchar <{\putsquarep}{\putsquarep%
   <\arrowtypea`\arrowtypeb`\arrowtypec`\arrowtyped;\width`\height>}}
\def\square{\@ifnextchar< {\squarep}{\squarep
   <\arrowtypea`\arrowtypeb`\arrowtypec`\arrowtyped;\width`\height>}}
\def\squarep<#1>[#2`#3`#4`#5;#6`#7`#8`#9]{{
\setsqparms[#1]
\diagram
\putsquarep<\arrowtypea`\arrowtypeb`\arrowtypec`
\arrowtyped;\width`\height>
(0,0)[#2`#3`#4`{#5};#6`#7`#8`{#9}]
\enddiagram
}}                                                 
\def\putptrianglep<#1>(#2,#3)[#4`#5`#6;#7`#8`#9]{{%
\settriparms[#1]%
\xpos=#2 \ypos=#3 \advance\ypos by \height
\puthmorphism(\xpos,\ypos)[#4`#5`{#7}]{\height}{\arrowtypea}a%
\putvmorphism(\xpos,\ypos)[`#6`{#8}]{\height}{\arrowtypeb}l%
\advance\xpos by\height
\putmorphism(\xpos,\ypos)(-1,-1)[``{#9}]{\height}{\arrowtypec}r%
}}

\def\putptriangle{\@ifnextchar <{\putptrianglep}{\putptrianglep
   <\arrowtypea`\arrowtypeb`\arrowtypec;\height>}}
\def\ptriangle{\@ifnextchar <{\ptrianglep}{\ptrianglep
   <\arrowtypea`\arrowtypeb`\arrowtypec;\height>}}
\def\ptrianglep<#1>[#2`#3`#4;#5`#6`#7]{{
\settriparms[#1]
\diagram
\putptrianglep<\arrowtypea`\arrowtypeb`
\arrowtypec;\height>
(0,0)[#2`#3`#4;#5`#6`{#7}]
\enddiagram
}}                                            

\def\putqtrianglep<#1>(#2,#3)[#4`#5`#6;#7`#8`#9]{{%
\settriparms[#1]%
\xpos=#2 \ypos=#3 \advance\ypos by\height
\puthmorphism(\xpos,\ypos)[#4`#5`{#7}]{\height}{\arrowtypea}a%
\putmorphism(\xpos,\ypos)(1,-1)[``{#8}]{\height}{\arrowtypeb}l%
\advance\xpos by\height
\putvmorphism(\xpos,\ypos)[`#6`{#9}]{\height}{\arrowtypec}r%
}}

\def\putqtriangle{\@ifnextchar <{\putqtrianglep}{\putqtrianglep
   <\arrowtypea`\arrowtypeb`\arrowtypec;\height>}}
\def\qtriangle{\@ifnextchar <{\qtrianglep}{\qtrianglep
   <\arrowtypea`\arrowtypeb`\arrowtypec;\height>}}
\def\qtrianglep<#1>[#2`#3`#4;#5`#6`#7]{{
\settriparms[#1]
\width=\height                                
\diagram
\putqtrianglep<\arrowtypea`\arrowtypeb`
\arrowtypec;\height>
(0,0)[#2`#3`#4;#5`#6`{#7}]
\enddiagram
}}

\def\putdtrianglep<#1>(#2,#3)[#4`#5`#6;#7`#8`#9]{{%
\settriparms[#1]%
\xpos=#2 \ypos=#3
\puthmorphism(\xpos,\ypos)[#5`#6`{#9}]{\height}{\arrowtypec}b%
\advance\xpos by \height \advance\ypos by\height
\putmorphism(\xpos,\ypos)(-1,-1)[``{#7}]{\height}{\arrowtypea}l%
\putvmorphism(\xpos,\ypos)[#4``{#8}]{\height}{\arrowtypeb}r%
}}

\def\putdtriangle{\@ifnextchar <{\putdtrianglep}{\putdtrianglep
   <\arrowtypea`\arrowtypeb`\arrowtypec;\height>}}
\def\dtriangle{\@ifnextchar <{\dtrianglep}{\dtrianglep
   <\arrowtypea`\arrowtypeb`\arrowtypec;\height>}}
\def\dtrianglep<#1>[#2`#3`#4;#5`#6`#7]{{
\settriparms[#1]
\width=\height                                
\diagram
\putdtrianglep<\arrowtypea`\arrowtypeb`
\arrowtypec;\height>
(0,0)[#2`#3`#4;#5`#6`{#7}]
\enddiagram
}}

\def\putbtrianglep<#1>(#2,#3)[#4`#5`#6;#7`#8`#9]{{%
\settriparms[#1]%
\xpos=#2 \ypos=#3
\puthmorphism(\xpos,\ypos)[#5`#6`{#9}]{\height}{\arrowtypec}b%
\advance\ypos by\height
\putmorphism(\xpos,\ypos)(1,-1)[``{#8}]{\height}{\arrowtypeb}r%
\putvmorphism(\xpos,\ypos)[#4``{#7}]{\height}{\arrowtypea}l%
}}

\def\putbtriangle{\@ifnextchar <{\putbtrianglep}{\putbtrianglep
   <\arrowtypea`\arrowtypeb`\arrowtypec;\height>}}
\def\btriangle{\@ifnextchar <{\btrianglep}{\btrianglep
   <\arrowtypea`\arrowtypeb`\arrowtypec;\height>}}
\def\btrianglep<#1>[#2`#3`#4;#5`#6`#7]{{
\settriparms[#1]
\width=\height                               
\diagram
\putbtrianglep<\arrowtypea`\arrowtypeb`
\arrowtypec;\height>
(0,0)[#2`#3`#4;#5`#6`{#7}]
\enddiagram
}}

\def\putAtrianglep<#1>(#2,#3)[#4`#5`#6;#7`#8`#9]{{%
\settriparms[#1]%
\xpos=#2 \ypos=#3 {\multiply \height by2
\puthmorphism(\xpos,\ypos)[#5`#6`{#9}]{\height}{\arrowtypec}b}%
\advance\xpos by\height \advance\ypos by\height
\putmorphism(\xpos,\ypos)(-1,-1)[#4``{#7}]{\height}{\arrowtypea}l%
\putmorphism(\xpos,\ypos)(1,-1)[``{#8}]{\height}{\arrowtypeb}r%
}}

\def\putAtriangle{\@ifnextchar <{\putAtrianglep}{\putAtrianglep
   <\arrowtypea`\arrowtypeb`\arrowtypec;\height>}}
\def\Atriangle{\@ifnextchar <{\Atrianglep}{\Atrianglep
   <\arrowtypea`\arrowtypeb`\arrowtypec;\height>}}
\def\Atrianglep<#1>[#2`#3`#4;#5`#6`#7]{{
\settriparms[#1]
\width=\height                                     
\diagram
\putAtrianglep<\arrowtypea`\arrowtypeb`
\arrowtypec;\height>
(0,0)[#2`#3`#4;#5`#6`{#7}]
\enddiagram
}}

\def\putAtrianglepairp<#1>(#2)[#3;#4`#5`#6`#7`#8]{{%
\settripairparms[#1]%
\setpos(#2)%
\settokens`#3`%
\puthmorphism(\xpos,\ypos)[\tokenb`\tokenc`{#7}]{\height}{\arrowtyped}b%
\advance\xpos by\height
\puthmorphism(\xpos,\ypos)[\phantom{\tokenc}`\tokend`{#8}]%
{\height}{\arrowtypee}b%
\advance\ypos by\height
\putmorphism(\xpos,\ypos)(-1,-1)[\tokena``{#4}]{\height}{\arrowtypea}l%
\putvmorphism(\xpos,\ypos)[``{#5}]{\height}{\arrowtypeb}m%
\putmorphism(\xpos,\ypos)(1,-1)[``{#6}]{\height}{\arrowtypec}r%
}}

\def\putAtrianglepair{\@ifnextchar <{\putAtrianglepairp}{\putAtrianglepairp%
   <\arrowtypea`\arrowtypeb`\arrowtypec`\arrowtyped`\arrowtypee;\height>}}
\def\Atrianglepair{\@ifnextchar <{\Atrianglepairp}{\Atrianglepairp%
   <\arrowtypea`\arrowtypeb`\arrowtypec`\arrowtyped`\arrowtypee;\height>}}

\def\Atrianglepairp<#1>[#2;#3`#4`#5`#6`#7]{{
\settripairparms[#1]
\settokens`#2`
\width=\height                                
\diagram
\putAtrianglepairp                            
<\arrowtypea`\arrowtypeb`\arrowtypec`
\arrowtyped`\arrowtypee;\height>
(0,0)[{#2};#3`#4`#5`#6`{#7}]
\enddiagram
}}

\def\putVtrianglep<#1>(#2,#3)[#4`#5`#6;#7`#8`#9]{{%
\settriparms[#1]%
\xpos=#2 \ypos=#3 \advance\ypos by\height {\multiply\height by2
\puthmorphism(\xpos,\ypos)[#4`#5`{#7}]{\height}{\arrowtypea}a}%
\putmorphism(\xpos,\ypos)(1,-1)[`#6`{#8}]{\height}{\arrowtypeb}l%
\advance\xpos by\height \advance\xpos by\height
\putmorphism(\xpos,\ypos)(-1,-1)[``{#9}]{\height}{\arrowtypec}r%
}}

\def\putVtriangle{\@ifnextchar <{\putVtrianglep}{\putVtrianglep
   <\arrowtypea`\arrowtypeb`\arrowtypec;\height>}}
\def\Vtriangle{\@ifnextchar <{\Vtrianglep}{\Vtrianglep
   <\arrowtypea`\arrowtypeb`\arrowtypec;\height>}}
\def\Vtrianglep<#1>[#2`#3`#4;#5`#6`#7]{{
\settriparms[#1]
\width=\height                                 
\diagram
\putVtrianglep<\arrowtypea`\arrowtypeb`
\arrowtypec;\height>
(0,0)[#2`#3`#4;#5`#6`{#7}]
\enddiagram
}}

\def\putVtrianglepairp<#1>(#2)[#3;#4`#5`#6`#7`#8]{{
\settripairparms[#1]%
\setpos(#2)%
\settokens`#3`%
\advance\ypos by\height
\putmorphism(\xpos,\ypos)(1,-1)[`\tokend`{#6}]{\height}{\arrowtypec}l%
\puthmorphism(\xpos,\ypos)[\tokena`\tokenb`{#4}]{\height}{\arrowtypea}a%
\advance\xpos by\height
\puthmorphism(\xpos,\ypos)[\phantom{\tokenb}`\tokenc`{#5}]%
{\height}{\arrowtypeb}a%
\putvmorphism(\xpos,\ypos)[``{#7}]{\height}{\arrowtyped}m%
\advance\xpos by\height
\putmorphism(\xpos,\ypos)(-1,-1)[``{#8}]{\height}{\arrowtypee}r%
}}

\def\putVtrianglepair{\@ifnextchar <{\putVtrianglepairp}{\putVtrianglepairp%
    <\arrowtypea`\arrowtypeb`\arrowtypec`\arrowtyped`\arrowtypee;\height>}}
\def\Vtrianglepair{\@ifnextchar <{\Vtrianglepairp}{\Vtrianglepairp%
    <\arrowtypea`\arrowtypeb`\arrowtypec`\arrowtyped`\arrowtypee;\height>}}
\def\Vtrianglepairp<#1>[#2;#3`#4`#5`#6`#7]{{
\settripairparms[#1]
\settokens`#2`
\diagram
\putVtrianglepairp                             
<\arrowtypea`\arrowtypeb`\arrowtypec`
\arrowtyped`\arrowtypee;\height>
(0,0)[{#2};#3`#4`#5`#6`{#7}]
\enddiagram
}}

\def\putCtrianglep<#1>(#2,#3)[#4`#5`#6;#7`#8`#9]{{%
\settriparms[#1]%
\xpos=#2 \ypos=#3 \advance\ypos by\height
\putmorphism(\xpos,\ypos)(1,-1)[``{#9}]{\height}{\arrowtypec}l%
\advance\xpos by\height \advance\ypos by\height
\putmorphism(\xpos,\ypos)(-1,-1)[#4`#5`{#7}]{\height}{\arrowtypea}l%
{\multiply\height by 2
\putvmorphism(\xpos,\ypos)[`#6`{#8}]{\height}{\arrowtypeb}r}%
}}

\def\putCtriangle{\@ifnextchar <{\putCtrianglep}{\putCtrianglep
    <\arrowtypea`\arrowtypeb`\arrowtypec;\height>}}
\def\Ctriangle{\@ifnextchar <{\Ctrianglep}{\Ctrianglep
    <\arrowtypea`\arrowtypeb`\arrowtypec;\height>}}
\def\Ctrianglep<#1>[#2`#3`#4;#5`#6`#7]{{
\settriparms[#1]
\width=\height                               
\diagram
\putCtrianglep<\arrowtypea`\arrowtypeb`
\arrowtypec;\height>
(0,0)[#2`#3`#4;#5`#6`{#7}]
\enddiagram
}}                                           
\def\putDtrianglep<#1>(#2,#3)[#4`#5`#6;#7`#8`#9]{{%
\settriparms[#1]%
\xpos=#2 \ypos=#3 \advance\xpos by\height \advance\ypos by\height
\putmorphism(\xpos,\ypos)(-1,-1)[``{#9}]{\height}{\arrowtypec}r%
\advance\xpos by-\height \advance\ypos by\height
\putmorphism(\xpos,\ypos)(1,-1)[`#5`{#8}]{\height}{\arrowtypeb}r%
{\multiply\height by 2
\putvmorphism(\xpos,\ypos)[#4`#6`{#7}]{\height}{\arrowtypea}l}%
}}

\def\putDtriangle{\@ifnextchar <{\putDtrianglep}{\putDtrianglep
    <\arrowtypea`\arrowtypeb`\arrowtypec;\height>}}
\def\Dtriangle{\@ifnextchar <{\Dtrianglep}{\Dtrianglep
   <\arrowtypea`\arrowtypeb`\arrowtypec;\height>}}
\def\Dtrianglep<#1>[#2`#3`#4;#5`#6`#7]{{
\settriparms[#1]
\width=\height                              
\diagram
\putDtrianglep<\arrowtypea`\arrowtypeb`
\arrowtypec;\height>
(0,0)[#2`#3`#4;#5`#6`{#7}]
\enddiagram
}}                                          
\def\setrecparms[#1`#2]{\width=#1 \height=#2}%

\def\recursep<#1`#2>[#3;#4`#5`#6`#7`#8]{{\m@th
\width=#1 \height=#2 \settokens`#3`
\settowidth{\tempdimen}{$\tokena$} \ifdim\tempdimen=0pt
  \savebox{\tempboxa}{\hbox{$\tokenb$}}%
  \savebox{\tempboxb}{\hbox{$\tokend$}}%
  \savebox{\tempboxc}{\hbox{$#6$}}%
\else
  \savebox{\tempboxa}{\hbox{$\hbox{$\tokena$}\times\hbox{$\tokenb$}$}}%
  \savebox{\tempboxb}{\hbox{$\hbox{$\tokena$}\times\hbox{$\tokend$}$}}%
  \savebox{\tempboxc}{\hbox{$\hbox{$\tokena$}\times\hbox{$#6$}$}}%
\fi \ypos=\height \divide\ypos by 2 \xpos=\ypos \advance\xpos by
\width \bfig
\putCtrianglep<-1`1`1;\ypos>(0,0)[`\tokenc`;#5`#6`{#7}]%
\puthmorphism(\ypos,0)[\tokend`\usebox{\tempboxb}`{#8}]{\width}{-1}b%
\puthmorphism(\ypos,\height)[\tokenb`\usebox{\tempboxa}`{#4}]{\width}{-1}a%
\advance\ypos by \width
\putvmorphism(\ypos,\height)[``\usebox{\tempboxc}]{\height}1r%
\efig }}

\def\recurse{\@ifnextchar <{\recursep}{\recursep<\width`\height>}}

\def\puttwohmorphisms(#1,#2)[#3`#4;#5`#6]#7#8#9{{%
\puthmorphism(#1,#2)[#3`#4`]{#7}0a \ypos=#2 \advance\ypos by 20
\puthmorphism(#1,\ypos)[\phantom{#3}`\phantom{#4}`#5]{#7}{#8}a
\advance\ypos by -40
\puthmorphism(#1,\ypos)[\phantom{#3}`\phantom{#4}`#6]{#7}{#9}b }}

\def\puttwovmorphisms(#1,#2)[#3`#4;#5`#6]#7#8#9{{%
\putvmorphism(#1,#2)[#3`#4`]{#7}0a \xpos=#1 \advance\xpos by -20
\putvmorphism(\xpos,#2)[\phantom{#3}`\phantom{#4}`#5]{#7}{#8}l
\advance\xpos by 40
\putvmorphism(\xpos,#2)[\phantom{#3}`\phantom{#4}`#6]{#7}{#9}r }}

\def\puthcoequalizer(#1)[#2`#3`#4;#5`#6`#7]#8#9{{%
\setpos(#1)%
\puttwohmorphisms(\xpos,\ypos)[#2`#3;#5`#6]{#8}11%
\advance\xpos by #8
\puthmorphism(\xpos,\ypos)[\phantom{#3}`#4`#7]{#8}1{#9} }}

\def\putvcoequalizer(#1)[#2`#3`#4;#5`#6`#7]#8#9{{%
\setpos(#1)%
\puttwovmorphisms(\xpos,\ypos)[#2`#3;#5`#6]{#8}11%
\advance\ypos by -#8
\putvmorphism(\xpos,\ypos)[\phantom{#3}`#4`#7]{#8}1{#9} }}

\def\putthreehmorphisms(#1)[#2`#3;#4`#5`#6]#7(#8)#9{{%
\setpos(#1) \settypes(#8)
\if a#9 %
     \vertsize{\tempcounta}{#5}%
     \vertsize{\tempcountb}{#6}%
     \ifnum \tempcounta<\tempcountb \tempcounta=\tempcountb \fi
\else
     \vertsize{\tempcounta}{#4}%
     \vertsize{\tempcountb}{#5}%
     \ifnum \tempcounta<\tempcountb \tempcounta=\tempcountb \fi
\fi \advance \tempcounta by 60
\puthmorphism(\xpos,\ypos)[#2`#3`#5]{#7}{\arrowtypeb}{#9}
\advance\ypos by \tempcounta
\puthmorphism(\xpos,\ypos)[\phantom{#2}`\phantom{#3}`#4]{#7}{\arrowtypea}{#9}
\advance\ypos by -\tempcounta \advance\ypos by -\tempcounta
\puthmorphism(\xpos,\ypos)[\phantom{#2}`\phantom{#3}`#6]{#7}{\arrowtypec}{#9}
}}

\def\setarrowtoks[#1`#2`#3`#4`#5`#6]{%
\def\toka{#1}
\def\tokb{#2}
\def\tokc{#3}
\def\tokd{#4}
\def\toke{#5}
\def\tokf{#6}
}
\def\hex{\@ifnextchar <{\hexp}{\hexp<1000`400>}}
\def\hexp<#1`#2>[#3`#4`#5`#6`#7`#8;#9]{%
\setarrowtoks[#9] \yext=#2 \advance \yext by #2 \xext=#1
\advance\xext by \yext \bfig
\putCtriangle<-1`0`1;#2>(0,0)[`#5`;\tokb``\tokd] \xext=#1
\yext=#2 \advance \yext by #2
\putsquare<1`0`0`1;\xext`\yext>(#2,0)[#3`#4`#7`#8;\toka```\tokf]
\advance \xext by #2
\putDtriangle<0`1`-1;#2>(\xext,0)[`#6`;`\tokc`\toke] \efig }

\makeatother

\input{tcilatex}

\begin{document}

\begin{frontmatter}

\title{Phase Transitions, Chaos and Joint Action in the Life Space Foam}
\author{Vladimir Ivancevic, Eugene Aidman, Leong Yen and Darryn Reid}
\address{Land Operations Division,
Defence Science \& Technology Organisation, AUSTRALIA}

\begin{abstract}

This paper extends our recently developed Life Space Foam (LSF)
model of motivated cognitive dynamics \cite{IA}. LSF uses adaptive
path integrals to generate Lewinian force--fields on smooth
manifolds, in order to characterize the dynamics of individual
goal--directed action. According to explanatory theories growing
in acceptance in cognitive neuroscience, one of the key properties
of this dynamics, capable of linking it to microscopic-level
cortical neurodynamics, is its meta-stability and the resulting
phase transitions. Our extended LSF model incorporates the notion
of phase transitions and complements it with embedded geometrical
chaos. To describe this LSF phase transition, a general
path--integral is used, along the corresponding LSF topology
change. As a result, our extended LSF model is able to rigorously
represent co-action by two or more actors in the common
LSF--manifold. The model yields substantial qualitative
differences in geometrical properties between bilateral and
multi-lateral co-action due to intrinsic chaotic coupling between
$n$ actors when $n\geq 3$.\bigbreak

\end{abstract}


\end{frontmatter}

\section{Introduction}

General stochastic dynamics, developed in a framework of Feynman
path integrals, have recently \cite{IA} been applied to Lewinian
field--theoretic psychodynamics \cite{Lewin51,Lewin97,Gold},
resulting in the development of a new concept of {life--space
foam} (LSF) as a natural medium for motivational (MD) and
cognitive (CD) psychodynamics. According to the LSF--formalism,
the classic Lewinian life space can be macroscopically represented
as a smooth manifold with steady force--fields and behavioral
paths, while at the microscopic level it is more realistically
represented as a collection of wildly fluctuating force--fields,
(loco)motion paths and local geometries (and topologies with
holes).

A set of least--action principles is used to model the smoothness
of global, macro--level LSF paths, fields and geometry, according
to the following prescription. The action $S[\Phi]$,
psycho--physical dimensions of $Energy\times Time=Effort$ and
depending on macroscopic paths, fields and geometries (commonly
denoted by an abstract field symbol $\Phi^i$) is defined as a
temporal integral from the {initial} time instant $t_{ini}$ to the
{final} time instant $t_{fin}$,
\begin{equation}
S[\Phi]=\int_{t_{ini}}^{t_{fin}}\mathfrak{L}[\Phi]\,dt,
\label{act}
\end{equation}%
with {Lagrangian density} given by
\begin{equation*}
\mathfrak{L}[\Phi]=\int
d^{n}x\,\mathcal{L}(\Phi_i,\partial_{x^j}\Phi^i),
\end{equation*}%
where the integral is taken over all $n$ coordinates $x^j=x^j(t)$
of the LSF, and $\partial_{x^j}\Phi^i$ are time and space partial
derivatives of the $\Phi^i-$variables over coordinates. The
standard {least action principle}
\begin{equation}
\delta S[\Phi]=0, \label{actPr}
\end{equation}
gives, in the form of the so--called Euler--Lagrangian equations,
a shortest (loco)motion path, an extreme force--field, and a
life--space geometry of minimal curvature (and without holes). In
this way, we effectively derive a {unique globally smooth
transition map}
\begin{equation}
F\;:\;INTENTION_{t_{ini}}\cone{} ACTION_{t_{fin}}, \label{unique}
\end{equation}
performed at a macroscopic (global) time--level from some initial
time $t_{ini}$ to the final time $t_{fin}$. In this way, we have
obtained macro--objects in the global LSF: a single path described
by Newtonian--like equation of motion, a single force--field
described by Maxwellian--like field equations, and a single
obstacle--free Riemannian geometry (with global topology without
holes).

To model the corresponding local, micro--level LSF structures of
rapidly fluctuating MD \& CD, an adaptive path integral is
formulated, defining a multi--phase and multi--path (multi--field
and multi--geometry) {transition amplitude} from the state of
$Intention$ to the state of $Action$,
\begin{equation}
\langle Action|Intention\rangle_{total} :=\put(0,0){\LARGE
$\int$}_{\rm
~~LSF}\put(-200,15){\large$\bf\Sigma$}\mathcal{D}[w\Phi]\,
{\mathrm e}^{\mathrm i S[\Phi]}, \label{pathInt1}
\end{equation}
where the Lebesgue integration is performed over all continuous
$\Phi^i_{con}=paths+fields+geometries$, while summation is
performed over all discrete processes and regional topologies
$\Phi^j_{dis}$. The symbolic differential $\mathcal{D}[w\Phi]$ in
the general path integral (\ref{pathInt}), represents an {adaptive
path measure}, defined as a weighted product
\begin{equation}
\mathcal{D}[w\Phi]=\lim_{N\to\infty}\prod_{s=1}^{N}w_sd\Phi
_{s}^i, \qquad ({i=1,...,n=con+dis}).\label{prod1}
\end{equation}
The adaptive path integral (\ref{pathInt1})--(\ref{prod1})
represents an $\infty-$dimensional neural network, with weights
$w$ updating by the general rule \cite{IA}
$$new\;value(t+1)\; =\; old\;value(t)\;+\; innovation(t).$$

On the other hand, it is well--known that phase transitions (PTs)
are phenomena which bring about {qualitative} physical changes at
the macroscopic level in presence of the same microscopic forces
acting among the constituents of a system. Their mathematical
description requires to translate into {quantitative} terms the
mentioned qualitative changes. The standard way of doing this is
to consider how the values of thermodynamic observables, obtained
in laboratory experiments, vary with temperature, or volume, or an
external field, and then to associate the experimentally observed
discontinuities at a PT to the appearance of some kind of
singularity entailing a loss of analyticity \cite{FP04}. Despite
the smoothness of the statistical
measures, after the {Yang--Lee theorem} \cite{YLthm} we know that in the $%
N\rightarrow \infty $ limit non--analytic behaviors of
thermodynamic functions are possible whenever the analyticity
radius in the complex
fugacity plane shrinks to zero, because this entails the loss of {%
uniform convergence} in $N$ (number of degrees of freedom) of any
sequence of real-valued thermodynamic functions, and all this
depends on the distribution of the zeros of the grand canonical
partition function. Also the other developments of the rigorous
theory of PTs (see, e.g., \cite{georgii,ruelleTD}), identify PTs
with the {loss of analyticity}.

Similarly, experimental findings and theoretical insights of
modern neuroscience converge on interpreting brain physiology
within the conceptual framework of nonlinear dynamics, operating
at the brink of criticality, which is achieved and maintained by
self-organization \cite{Werner}. In this approach, dynamical
patterning of both brain activity and corresponding behaviors is
examined in order to develop models of how brain and behavioral
events are coordinated. Growing evidence supports the assumption
that the linkages between events at microscopic level of neuronal
assemblies and those at macroscopic behavioral levels are better
explained as based on shared dynamics -- not on any ontological
priority \cite{Kelso92,Erlhagen}. This dynamics is characterized
by meta-stability and phase transitions in self-organized
criticality \cite{Bak}, both at the level of neuronal assemblies
\cite{Fre05,Werner}, functional connectivity of the human brain
\cite{Chialvo} and corresponding behavior patterns
\cite{Kelso92,Gro08}. A key feature of this approach is that
phenomenological laws at the behavioral level can be connected to
a field-theoretical description of cortical dynamics
\cite{Kelso92}. Dynamic Field Theory (DFT) \cite{Amari} extends
this approach by developing field-theoretic representations of
both behavior and its environment \cite{Schoner}, thus building on
a long established, albeit metaphorical, tradition of behavioral
force-field analysis \cite{Lewin51,Lewin97,Gold}. Our
LSF--formalism can be seen as a further extension of DFT.

Regarding brain modelling: classical physics has provided a strong
foundation for understanding brain function through measuring
brain activity, modelling the functional connectivity of networks
of neurons with algebraic matrices, and modelling the dynamics of
neurons and neural populations with sets of coupled differential
equations \cite{Fre75,Fre00}. Various tools from classical physics
enabled recognition and documentation of aspects of the physical
states of the brain; the structures and dynamics of neurons, the
operations of membranes and organelles that generate and channel
electric currents; and the molecular and ionic carriers that
implement the neural machineries of electrogenesis and learning.
They support description of brain functions at several levels of
complexity through measuring neural activity in the brains of
animal and human subjects engaged in behavioral exchanges with
their environments. One of the key properties of brain dynamics
are the coordinated oscillations of populations of neurons that
change rapidly in concert with changes in the environment
\cite{FreVit06}.

Also, most experimental neurobiologists and neural theorists have
focused on sensorimotor functions and their adaptations through
various forms of learning and memory. Reliance has been placed on
measurements of the rates and intervals of trains of action
potentials of small numbers of neurons that are tuned to
perceptual invariances and modelling neural interactions with
discrete networks of simulated neurons. These and related studies
have given a vivid picture of the cortex as a mosaic of modules,
each of which performs a sensory or motor function; they have not
given a picture of comparable clarity of the integration of
modules (see \cite{FreVit06} and references therein).

The EEG analysis performed on rabbits and cats trained to
discriminate conditioned stimuli in the various modalities (with
EEG-recordings collected from high-density electrode arrays fixed
on the epidural surfaces of primary sensory and limbic areas) has
shown that cortical activity does not change continuously with
time but by multiple spatial patterns in sequences during each
perceptual action that resemble cinematographic frames on multiple
screens \cite{FreVit06}. The carrier waves of the patterned
activity in frames have come in at least two ranges identified
with beta (12-30 Hz) and gamma (30-80 Hz) oscillations. The abrupt
change in dynamical state with each new frame, proposed to be
formed by a phase transition \cite{Fre04}, has not been
describable either with classic integro-differential equations, or
the algebras of neural networks. The initiation and maintenance of
shared oscillations by this phase transition requires rapid
communication among neurons. Several alternative mechanisms have
been proposed as the agency for widespread synchrony. These are
based in the dendritic loop current as the chief agent for
intracellular communication and the axonal action potential as the
chief agent for intercellular communication \cite{FreVit06}.

According to \cite{FreVit06}, {many-body quantum field theory}
appears to be the only existing theoretical tool capable to
explain the dynamic origin of long-range correlations, their rapid
and efficient formation and dissolution, their interim stability
in ground states, the multiplicity of coexisting and possibly
non-interfering ground states, their degree of ordering, and their
rich textures relating to sensory and motor facets of behaviors.
It is historical fact that many-body quantum field theory has been
devised and constructed in past decades exactly to understand
features like ordered pattern formation and phase transitions in
condensed matter physics that could not be understood in classical
physics, similar to those in the brain.

Communication by propagating action potentials imposes
distance-dependent delays in the onset of re-synchronization
during a phase transition over an area of cortex. The delays are
measurable as brief but distance-dependent phase lags at the
various frequencies of oscillation \cite{Fre04}. However, the
length of most axons in cortex is a small fraction of observed
distances of long-range correlation, with the requirement for
synaptic renewal at each successive relay. These {long-range
correlations} are maintained despite continuous variations in
transmission frequencies that are apparent in aperiodic `chaotic'
oscillations.

Some researchers have sought to explain zero-lag correlations with
processes other than axo-dendritic synaptic transmission, stating
that both electric fields and magnetic fields accompany neural
loop currents. However, the electric potential gradients of the
EEG have been shown by \cite{FreBai} to be inadequate {in vivo} to
account for the long-range of the observed coherent activity,
largely owing to the shunting action of glia that reduce the
fraction of extracellular dendritic current penetrating adjacent
neurons and minimize ephaptic cross-talk among cortical neurons.

In this paper, to describe the LSF--phase--transitions, with
embedded chaos, we use our adaptive path--integral (\ref{pathInt})
along the corresponding LSF--topology--change:
$$\langle {\rm phase~out}\,|\,{\rm
phase~in}\rangle\, :=\,\put(0,0){\LARGE $\int$}_{\rm
~~topology-change}\put(-610,20){\large$\bf\Sigma$}\mathcal{D}[w\Phi]\,
{\mathrm e}^{\mathrm i S[\Phi]}.
$$

This paper extends the earlier establishment of the LSF model by
introducing the study of chaos and phase transitions within this
framework. This development is motivated -- as was the original
LSF model -- by the potential for an improved theoretical basis
for studying brain physiology, and thereby also for overcoming
shortfalls in current artificial neural network function
representation and approximation technologies.

\section{Geometrical Chaos and Topological Phase Transitions}

In this section we extend the LSF--formalism to incorporate
geometrical chaos and associated topological phase transitions.

It is well--known that on the basis of the {ergodic hypothesis},
statistical mechanics describes the physics of many-degrees of
freedom systems by
replacing {time averages} of the relevant observables with {%
ensemble averages}. Therefore, instead of using statistical
ensembles, we can investigate the Hamiltonian (microscopic)
dynamics of a system undergoing a phase transition. The reason for
tackling dynamics is twofold. First, there are observables, like
Lyapunov exponents, that are intrinsically dynamical. Second, the
geometrization of Hamiltonian dynamics in terms of Riemannian
geometry provides new observables and, in general, an interesting
framework to investigate the phenomenon of phase transitions
\cite{CCC97,Pet07}. The geometrical formulation of the {dynamics
of conservative systems} \cite{AbrahamMeh} was first used by
\cite{Krylov} in his studies on the dynamical foundations of
statistical mechanics and subsequently became a standard tool to
study abstract systems in ergodic theory.

The simplest, mechanical--like LSF--action in the individual's
LSF--manifold $\Sigma$ has a Riemannian locomotion form \cite{IA}
\begin{equation}
S[q]={1\over2}\int_{t_{ini}}^{t_{fin}}[a_{ij}
\,\dot{q}^{i}\dot{q}^{j}-V(q)]\,dt,\qquad \text{(summation
convention is assumed)} \label{locAct}
\end{equation}
where $a_{ij}$ is the `material' metric tensor that generates the
total `kinetic energy' of {cognitive} (loco)motions defined by
their configuration coordinates $q^i$ and velocities
$\dot{q}^{i}$, with the {motivational potential energy} $V(q)$ and
the standard Hamiltonian
\begin{equation}
H(p,q)=\sum_{i=1}^{N}\frac{1}{2}p_{i}^{2}+V(q), \label{Ham}
\end{equation}
where $p_{i}$ are the canonical (loco)motion momenta.

Dynamics of $N$ DOF mechanical--like systems with action
(\ref{locAct}) and Hamiltonian (\ref{Ham}) are commonly given by
the set of {geodesic equations} \cite{GaneshSprBig,GaneshADG}
\begin{equation}
\frac{d^{2}q^{i}}{ds^{2}}+\Gamma _{jk}^{i}\frac{dq^{j}}{ds}\frac{dq^{k}}{ds}%
=0,  \label{geod-mot}
\end{equation}%
where $\Gamma _{jk}^{i}$ are the Christoffel symbols of the affine
Levi--Civita connection of the Riemannian LSF--manifold $\Sigma$.

Alternatively, a description of the extrema of the Hamilton's
action (\ref{locAct}) can be obtained using the {Eisenhart metric}
\cite{Eisenhart} on an enlarged LSF space-time manifold (given by
$\{q^{0}\equiv t,q^{1},\ldots ,q^{N}\}$ plus one real coordinate
$q^{N+1}$), whose arc--length is
\begin{equation}
ds^{2}=-2V(q)(dq^{0})^{2}+a_{ij}dq^{i}dq^{j}+2dq^{0}dq^{N+1}.  \label{ds2E}
\end{equation}%
The manifold has a {Lorentzian structure} \cite{Pet07} and the
dynamical trajectories are those geodesics satisfying the
condition $ds^{2}=Cdt^{2}$, where $C$ is a positive constant. In
this geometrical framework, the instability of the trajectories is
the instability of the geodesics, and it is completely determined
by the curvature properties of the LSF--manifold $\Sigma$
according to the {Jacobi equation} of geodesic deviation
\cite{GaneshSprBig,GaneshADG}
\begin{equation}
\frac{D^{2}J^{i}}{ds^{2}}+R_{~jkm}^{i}\frac{dq^{j}}{ds}J^{k}\frac{dq^{m}}{ds}%
=0,  \label{eqJ}
\end{equation}%
whose solution $J$, usually called {Jacobi variation field},
locally measures the distance between nearby geodesics; $D/ds$
stands for the {covariant derivative} along a geodesic and
$R_{~jkm}^{i}$ are the components of the {Riemann curvature
tensor} of the LSF--manifold $\Sigma$.

Using the Eisenhart metric (\ref{ds2E}), the relevant part of the
Jacobi equation (\ref{eqJ}) is given by the {tangent dynamics
equation} \cite{pre96,CCC97}
\begin{equation}
\frac{d^{2}J^{i}}{dt^{2}}+R_{~0k0}^{i}J^{k}=0,\qquad (i=1,\dots
,N), \label{eqdintang}
\end{equation}%
where the only non-vanishing components of the curvature tensor of
the LSF--manifold $\Sigma$ are
\begin{equation*}
R_{~0k0}^{i}=\partial ^{2}V/\partial q^{i}\partial q^{j}.
\end{equation*}

The tangent dynamics equation (\ref{eqdintang}) is commonly used
to define {Lyapunov exponents} in dynamical systems given by the
Riemannian action (\ref{locAct}) and Hamiltonian (\ref{Ham}),
using the formula \cite{physrep}
\begin{equation}
\lambda _{1}=\lim_{t\rightarrow \infty }1/2t\log (\Sigma
_{i=1}^{N}[J_{i}^{2}(t)+J_{i}^{2}(t)]/\Sigma
_{i=1}^{N}[J_{i}^{2}(0)+J_{i}^{2}(0)]). \label{Lyap1}
\end{equation}
Lyapunov exponents measure the {strength of dynamical chaos}.

Now, to relate these results to topological phase transitions
within the LSF--manifold $\Sigma$, recall that any two
high--dimensional manifolds $\Sigma _{v}$ and $\Sigma _{v^{\prime
}}$ have the same topology if they can be continuously and
differentiably deformed into one another, that is if they are
diffeomorphic. Thus by {topology change} the `loss of
diffeomorphicity is meant \cite{Pet07}. In this respect, the
so--called {topological theorem} \cite{FP04} says that
non--analyticity is the `shadow' of a more fundamental phenomenon
occurring in the system's configuration manifold (in our case the
LSF--manifold): a topology change within the family of
equipotential hypersurfaces
\begin{equation*}
\Sigma _{v}=\{(q^{1},\dots ,q^{N})\in \mathbb{R}^{N}|\
V(q^{1},\dots ,q^{N})=v\},
\end{equation*}%
where $V$ and $q^{i}$ are the microscopic interaction potential
and coordinates respectively. This topological approach to PTs
stems from the numerical study of the dynamical counterpart of
phase transitions, and precisely from the observation of
discontinuous or cuspy patterns displayed by the largest Lyapunov
exponent $\lambda _{1}$ at the {transition energy} \cite{physrep}.
Lyapunov exponents cannot be measured in laboratory experiments,
at variance with thermodynamic observables, thus, being genuine
dynamical observables they are only be estimated in numerical
simulations of the microscopic dynamics. If there are critical points of $V$ in configuration space, that is points $%
q_{c}=[{\overline{q}}_{1},\dots ,{\overline{q}}_{N}]$ such that
$\left. \nabla V(q)\right\vert _{q=q_{c}}=0$, according to the
{Morse Lemma} \cite{hirsch}, in the neighborhood of any {critical
point} $q_{c}$
there always exists a coordinate system ${\tilde{q}}(t)=[{\tilde{q}}%
^{1}(t),..,{\tilde{q}}^{N}(t)]$ for which
\begin{equation}
V({\tilde{q}})=V(q_{c})-{\tilde{q}}_{1}^{2}-\dots -{\tilde{q}}_{k}^{2}+{%
\tilde{q}}_{k+1}^{2}+\dots +{\tilde{q}}_{N}^{2},  \label{morsechart}
\end{equation}%
where $k$ is the {index of the critical point}, i.e., the number
of negative eigenvalues of the Hessian of the potential energy
$V$. In the neighborhood of a critical point of the LSF--manifold
$\Sigma$, (\ref{morsechart}) yields
\begin{equation*}
\partial ^{2}V/\partial q^{i}\partial q^{j}=\pm \delta _{ij},
\end{equation*}%
which gives $k$ unstable directions which contribute to the
exponential growth of the norm of the tangent vector $J$
\cite{physrep}.

This means that the strength of dynamical chaos within the
individual's LSF--manifold $\Sigma$, measured by the largest
Lyapunov exponent $\lambda _{1}$ given by (\ref{Lyap1}), is
affected by the existence of critical points $q_{c}$ of the
potential energy $V(q)$. However, as $V(q)$ is bounded below, it
is a good {Morse function}, with no vanishing eigenvalues of its
Hessian matrix. According to {Morse theory} \cite{hirsch},
 the existence of critical points of $V$ is
associated with topology changes of the hypersurfaces $\{\Sigma
_{v}\}_{v\in \mathbb{R}}$.

More precisely, let $V_{N}(q_{1},\dots ,q_{N}):R^{N}\rightarrow R$, be a
smooth, bounded from below, finite-range and confining potential\footnote{%
These requirements for $V$ are fulfilled by standard interatomic and
intermolecular interaction potentials, as well as by classical spin
potentials.}. Denote by $\Sigma _{v}=V^{-1}(v)$, $v\in R$, its level sets,
or equipotential hypersurfaces, in the LSF--manifold $\Sigma$. Then let $\bar{v}%
=v/N $ be the potential energy per degree of freedom. If there exists $N_{0}$%
, and if for any pair of values $\bar{v}$ and $\bar{v}^{\prime }$ belonging
to a given interval $I_{\bar{v}}=[\bar{v}_{0},\bar{v}_{1}]$ and for any $%
N>N_{0} $ then the sequence of the Helmoltz free energies $\{F_{N}(\beta
)\}_{N\in \mathbb{N}}$ -- where $\beta =1/T$ ($T$ is the temperature) and $%
\beta \in I_{\beta }=(\beta (\bar{v}_{0}),\beta (\bar{v}_{1}))$ -- is
uniformly convergent at least in $C^{2}(I_{\beta })$ [the space of twice
differentiable functions in the interval $I_{\beta }$], so that $%
\lim_{N\rightarrow \infty }F_{N}\in C^{2}(I_{\beta })$ and neither
first nor second order phase transitions can occur in the
(inverse) temperature interval $(\beta (\bar{v}_{0}),\beta
(\bar{v}_{1}))$, where the inverse temperature is defined as
\cite{FP04,Pet07}
\begin{equation*}
\beta (\bar{v})=\partial S_{N}^{(-)}(\bar{v})/\partial \bar{v},\qquad \text{%
while\qquad }S_{N}^{(-)}(\bar{v})=N^{-1}\log \int_{V(q)\leq \bar{v}N}\ d^{N}q
\end{equation*}%
is one of the possible definitions of the microcanonical
configurational entropy. The intensive variable $\bar{v}$ has been
introduced to ease the comparison between quantities computed at
different $N$-values.

This theorem means that a topology change of the $\{\Sigma
_{v}\}_{v\in \mathbb{R}}$ at some $v_{c}$ is a {necessary}
condition for a phase transition to take place at the
corresponding energy value. The topology changes implied here are
those described within the framework of Morse theory through
`attachment of handles' \cite{hirsch} to the LSF--manifold
$\Sigma$.

In the LSF path--integral language \cite{IA}, we can say that
suitable topology changes of equipotential submanifolds of the
individual's LSF--manifold $\Sigma$ can entail thermodynamic--like
phase transitions \cite{Haken1,Haken2,Haken3}, according to the
general formula:
$$\langle {\rm phase~out}\,|\,{\rm
phase~in}\rangle\,:=\,\put(0,0){\LARGE $\int$}_{\rm
~~topology-change}\put(-610,20){\large$\bf\Sigma$}\mathcal{D}[w\Phi]\,
{\mathrm e}^{\mathrm i S[\Phi]}.
$$
The statistical behavior of the LSF--(loco)motion system
(\ref{locAct}) with the standard Hamiltonian (\ref{Ham}) is
encompassed, in the canonical ensemble, by its
{partition function}, given by the phase--space path integral \cite{GaneshADG}%
\begin{equation}
Z_{N}=\put(0,0){\LARGE $\int$}_{\rm
~~top-ch}\put(-300,20){\large$\bf\Sigma$}\mathcal{D}[p]\mathcal{D}[q]\exp \left\{ \mathrm{i}\int_{t}^{t{%
^{\prime }}}[p\dot{q}-H(p,q)]\,d\tau \right\} , \label{PI1}
\end{equation}%
where we have used the shorthand notation
\begin{equation*}
\put(0,0){\LARGE $\int$}_{\rm
~~top-ch}\put(-300,20){\large$\bf\Sigma$}\mathcal{D}[p]\mathcal{D}[q]\equiv
\int \prod_{\tau }\frac{dq(\tau )dp(\tau )}{2\pi }.
\end{equation*}%
The phase--space path integral (\ref{PI1}) can be calculated as
the partition function \cite{FPS00},
\begin{eqnarray}
Z_{N}(\beta ) &=&\int \prod_{i=1}^{N}dp_{i}dq^{i}{\rm e}^{-\beta
H(p,q)}=\left( \frac{\pi }{\beta }\right) ^{\frac{N}{2}}\!\!\int
\prod_{i=1}^{N}dq^{i}{\rm e}^{-\beta V(q)}  \notag \\
&=&\left( \frac{\pi }{\beta }\right)
^{\frac{N}{2}}\int_{0}^{\infty }dv\,{\rm e}^{-\beta v}\int_{\Sigma
_{v}}\frac{d\sigma }{\Vert \nabla V\Vert }\!\!,, \label{zeta}
\end{eqnarray}%
where the last term is written using the so--called {co--area formula} \cite%
{federer}, and $v$ labels the equipotential hypersurfaces $\Sigma
_{v}$ of the LSF--manifold $\Sigma$,
\begin{equation*}
\Sigma _{v}=\{(q^{1},\dots ,q^{N})\in \mathbb{R}^{N}|V(q^{1},\dots
,q^{N})=v\}.
\end{equation*}%
Equation (\ref{zeta}) shows that the relevant statistical
information is contained in the canonical configurational
partition function
\begin{equation*}
Z_{N}^{C}=\int \prod dq^{i}V(q){\rm e}^{-\beta V(q)}.
\end{equation*}%
Note that $Z_{N}^{C}$ is decomposed, in the last term of
(\ref{zeta}), into an infinite summation of geometric integrals,
\begin{equation*}
\int_{\Sigma _{v}}d\sigma \,/\Vert \nabla V\Vert ,
\end{equation*}%
defined on the $\{\Sigma _{v}\}_{v\in \mathbb{R}}$. Once the microscopic
interaction potential $V(q)$ is given, the configuration space of the system
is automatically foliated into the family $\{\Sigma _{v}\}_{v\in \mathbb{R}}$
of these equipotential hypersurfaces. Now, from standard statistical
mechanical arguments we know that, at any given value of the inverse
temperature $\beta $, the larger the number $N$, the closer to $%
\Sigma _{v}\equiv \Sigma _{u_{\beta }}$ are the microstates that
significantly contribute to the averages, computed through
$Z_{N}(\beta )$, of thermodynamic observables. The hypersurface
$\Sigma _{u_{\beta }}$ is the one associated with
\begin{equation*}
u_{\beta }=(Z_{N}^{C})^{-1}\int \prod dq^{i}V(q){\rm e}^{-\beta
V(q)},
\end{equation*}%
the average potential energy computed at a given $\beta $. Thus, at any $%
\beta $, if $N$ is very large the effective support of the
canonical measure shrinks very close to a single $\Sigma
_{v}=\Sigma _{u_{\beta }}$. Hence, the basic origin of a phase
transition lies in a suitable topology change of the $\{\Sigma
_{v}\}$, occurring at some $v_{c}$ \cite{FPS00}. This topology
change induces the singular behavior of the thermodynamic
observables at a phase transition. It is conjectured that the
counterpart of a phase transition is a breaking of
diffeomorphicity among the surfaces $\Sigma _{v}$, it is
appropriate to choose a {diffeomorphism invariant} to probe if and
how the topology of the $\Sigma _{v}$ changes as a function of
$v$. Fortunately, such a topological invariant exists, the {Euler
characteristic} of the LSF--manifold $\Sigma$, defined by
\cite{GaneshSprBig,GaneshADG}
\begin{equation}
\chi (\Sigma )=\sum_{k=0}^{N}(-1)^{k}b_{k}(\Sigma ),  \label{chi}
\end{equation}%
where the {Betti numbers} $b_{k}(\Sigma )$ are diffeomorphism invariants.\footnote{%
The Betti numbers $b_{k}$ are the dimensions of the de Rham's
cohomology vector spaces $H^{k}(\Sigma ;\mathbb{R})$ (therefore
the $b_{k}$ are integers).} This homological formula can be
simplified by the use of the {Gauss--Bonnet--Hopf theorem}, that
relates $\chi (\Sigma )$ with the total {Gauss--Kronecker
curvature} $K_{G}$ of the LSF--manifold $\Sigma$
\begin{equation}
\chi (\Sigma )=\int_{\Sigma }K_{G}\,d\sigma, \label{gaussbonnet}
\end{equation}%
where
\begin{equation*}
d\sigma =\sqrt{det(a)}dx^{1}dx^{2}\cdots dx^{n}
\end{equation*}%
is the invariant volume measure of the LSF--manifold $\Sigma$ and
$a$ is the determinant of the LSF metric tensor $a_{ij}$. For
technical details of this topological approach, see \cite{thorpe}.

The domain of validity of the `quantum' is not restricted to the
microscopic world \cite{Ume93}. There are macroscopic features of
classically behaving systems, which cannot be explained without
recourse to the quantum dynamics. This field theoretic model leads
to the view of the phase transition as a condensation that is
comparable to the formation of fog and rain drops from water
vapor, and that might serve to model both the gamma and beta phase
transitions. According to such a model, the production of activity
with long-range correlation in the brain takes place through the
mechanism of spontaneous breakdown of symmetry (SBS), which has
for decades been shown to describe long-range correlation in
condensed matter physics. The adoption of such a field theoretic
approach enables modelling of the whole cerebral hemisphere and
its hierarchy of components down to the atomic level as a fully
integrated macroscopic quantum system, namely as a macroscopic
system which is a quantum system not in the trivial sense that it
is made, like all existing matter, by quantum components such as
atoms and molecules, but in the sense that some of its macroscopic
properties can best be described with recourse to quantum dynamics
(see \cite{FreVit06} and references therein).

Phase transitions can also be associated with autonomous robot
competence levels, as informal specifications of desired classes
of behaviors for robots over all environments they will encounter,
as described by Brooks' subsumption architecture approach
\cite{BrooksLayered,BrooksWalk,BrooksElephants}. The distributed
network of augmented finite--state machines can exist in different
phases or modalities of their state--space variables, which
determine the systems intrinsic behavior. The phase transition
represented by this approach is triggered by either internal (a
set--point) or external (a command) control stimuli, such as a
command to transition from a sleep mode to awake mode, or walking
to running.

\section{Modelling Human Joint Action}

Cognitive neuroscience investigations, including fMRI studies of
human co-action, suggest that cognitive and neural processes
supporting co-action include joint attention, action observation,
task sharing, and action coordination
\cite{Fogassi,Knoblich,Newman,Sebanz}. For example, when two
actors are given a joint control task (e.g., tracking a moving
target on screen) and potentially conflicting controls (e.g., one
person in charge of acceleration, the other -- deceleration),
their joint performance depends on how well they can anticipate
each other's actions. In particular, better coordination is
achieved when individuals receive real-time feedback about the
timing of each other's actions \cite{Sebanz}.

To model the dynamics of the two--actor joint action, we propose
to associate each of the actors with an $n-$dimensional Riemannian
LSF--manifold $\Sigma$, that is a set of their own time dependent
trajectories, $\Sigma_{\alpha }=\{x^{i}(t_{i})\}$ and
$\Sigma_{\beta }=\{y^{j}(t_{j})\}$, respectively. Their associated
tangent bundles contain their individual $n$D (loco)motion
velocities, $T\Sigma_{\alpha }=\{\dot{x}^{i}(t_{i})=dx^{i}/dt_{i}\}$ and $%
T\Sigma_{\beta }=\{\dot{y}^{j}(t_{j})=dy^{j}/dt_{j}\}.$ Further,
following the general formalism of \cite{IA}, outlined in the
introduction, we use the modelling machinery consisting of: (i)
Adaptive joint action at the top--master level, describing the
externally--appearing deterministic, continuous and smooth
dynamics, and (ii) Corresponding adaptive path integral
(\ref{pathInt}) at the bottom--slave level, describing a wildly
fluctuating dynamics including both continuous trajectories and
Markov chains. This lower--level joint dynamics can be further
discretized into a partition function of the corresponding
statistical dynamics.

In particular, by extending and adapting classical Wheeler--Feynman {%
action--at--a--distance electrodynamics} \cite{FW} and applying it
to human co--action, we propose a two--term joint action:
\begin{eqnarray}
A[x,y;t_{i},t_{j}] &=&\frac{1}{2}\int_{t_{i}}\int_{t_{j}}\alpha
_{i}\beta
_{j}\,\delta (I_{ij}^{2})\,\,\dot{x}^{i}(t_{i})\,\dot{y}^{j}(t_{j})\,%
\,dt_{i}dt_{j}+{\frac{1}{2}}\int_{t}g_{ij}\,\dot{x}^{i}(t)\dot{x}^{j}(t)\,dt
\notag \\
\text{with\qquad }I_{ij}^{2} &=&\left[
x^{i}(t_{i})-y^{j}(t_{j})\right] ^{2},\qquad \text{where \  \
}IN\leq t_{i},t_{j},t\leq OUT.\hspace{2cm} \label{Fey1}
\end{eqnarray}

The first term in (\ref{Fey1}) represents {potential energy of the
cognitive/motivational interaction} between the two agents $\alpha
_{i}$ and
$\beta _{j}$.\footnote{%
Although, formally, this term contains cognitive velocities, it
still represents `potential energy' from the physical point of
view.} It is a double integral over a delta function of the square
of interval $I^{2}$ between two points on the paths in their
Life--Spaces; thus, interaction occurs only when this interval,
representing the motivational cognitive distance between the two
agents, vanishes. Note that the cognitive (loco)motions of the two
agents $\alpha _{i}[x^{i}(t_{i})]$ and $\beta _{j}[y^{j}(t_{j})]$,
generally occur at different times $t_{i}$ and $t_{j}$\ unless
$t_{i}=t_{j},$ when {cognitive synchronization} occurs.

The second term in (\ref{Fey1}) represents {kinetic energy of the
physical interaction}. Namely, when the cognitive synchronization
in the first term takes place, the second term of physical kinetic
energy is activated in the common manifold, which is one of the
agents' Life Spaces, say $\Sigma_{\alpha }=\{x^{i}(t_{i})\}$.

Conversely, if we have a need to represent coaction of three actors, say $\alpha _{i}$, $%
\beta _{j}$ and $\gamma _{k}$ (e.g., $\alpha _{i}$ in charge of
acceleration, $\beta _{j}$ -- deceleration and $\gamma _{k}-$
steering), we can associate each of them with an $n$D Riemannian
Life--Space manifold,
$\Sigma_{\alpha }=\{x^{i}(t_{i})\}$, $\Sigma_{\beta }=\{y^{j}(t_{j})\}$, and $%
\Sigma_{\gamma }=\{z^{k}(t_{k})\},$ respectively, with the
corresponding tangent
bundles containing their individual (loco)motion velocities, $T\Sigma_{\alpha }=\{%
\dot{x}^{i}(t_{i})=dx^{i}/dt_{i}\},T\Sigma_{\beta }=\{\dot{y}%
^{j}(t_{j})=dy^{j}/dt_{j}\}$ and $T\Sigma_\gamma=\{\dot{z}^{k}(t_{k})=dz^{k}/dt_{k}%
\}.$ Then, instead of (\ref{Fey1}) we have
\begin{eqnarray}
&&A[t_{i},t_{j},t_{k};t] =\frac{1}{2}\int_{t_{i}}\int_{t_{j}}\int_{t_{k}}%
\alpha _{i}(t_{i})\beta _{j}\,(t_{j})\,\gamma _{k}\,(t_{k})\delta
(I_{ijk}^{2})\,\,\dot{x}^{i}(t_{i})\,\dot{y}^{j}(t_{j})\,\dot{z}%
^{k}(t_{k})\,dt_{i}dt_{j}dt_{k} \notag  \label{Fey11} \\
&&+~{\frac{1}{2}}\int_{t}W_{rs}^{M}(t,q,\dot{q})\,\dot{q}^{r}\dot{q}%
^{s}\,dt,\,\qquad \text{where }IN \leq t_{i},t_{j},t_{k},t\leq OUT
\label{Fey2} \\
&&\text{with}\qquad I_{ijk}^{2}
=[x^{i}(t_{i})-y^{j}(t_{j})]^{2}+[y^{j}(t_{j})-z^{k}(t_{k})]^{2}+[z^{k}(t_{k})-x^{i}(t_{i})]^{2},
\notag  \label{Fey3}
\end{eqnarray}

Due to an {intrinsic chaotic coupling}, the three--actor (or,
$n-$actor, $n>3$) joint action (\ref{Fey2}) has a considerably
more complicated geometrical structure then the bilateral
co--action (\ref{Fey1}).\footnote{Recall that the {necessary}
condition for chaos in continuous temporal or spatio-temporal
systems is to have
three variables with nonlinear couplings between them.} It actually happens in the common $3n$D {%
Finsler manifold} $\Sigma_{J}=\Sigma_{\alpha }\cup \Sigma_{\beta
}\cup \Sigma_\gamma$,
parameterized by the local joint coordinates dependent on the common time $t$%
. That is, $\Sigma_{J}=\{q^{r}(t),\,r=1,...,3n\}.$ Geometry of the
joint manifold
$\Sigma_{J}$ is defined by the {Finsler metric function} $%
ds=F(q^{r},dq^{r}),$ defined by
\begin{equation}
F^{2}(q,\dot{q})=g_{rs}(q,\dot{q})\dot{q}^{r}\dot{q}^{s},\qquad
\text{(where }g_{rs}\text{ is the Riemann metric tensor)}
\label{Fins1}
\end{equation}%
\ and the {Finsler tensor} $C_{rst}(q,\dot{q}),$ defined by (see \cite%
{GaneshSprBig,GaneshADG})%
\begin{equation}
C_{rst}(q,\dot{q})=\frac{1}{4}\frac{\partial
^{3}F^{2}(q,\dot{q})}{\partial
\dot{q}^{r}\partial \dot{q}^{s}\partial \dot{q}^{t}}=\frac{1}{2}\frac{%
\partial g_{rs}}{\partial \dot{q}^{r}\partial \dot{q}^{s}}.  \label{Fins2}
\end{equation}%
From the Finsler definitions (\ref{Fins1})--(\ref{Fins2}), it
follows that the partial interaction manifolds, $\Sigma_{\alpha
}\cup \Sigma_{\beta },$ $\Sigma_{\beta }\cup \Sigma_{y}$ and
$\Sigma_{\alpha }\cup \Sigma_{y}$, have Riemannian structures with
the
corresponding interaction kinetic energies, $$T_{\alpha \beta }=\frac{1}{2}%
g_{ij}\dot{x}^{i}\dot{y}^{j},\qquad T_{\alpha \gamma }=\frac{1}{2}g_{ik}\dot{x}%
^{i}\dot{z}^{k},\qquad T_{\beta \gamma }=\frac{1}{2}g_{jk}\dot{y}^{j}\dot{z}%
^{k}.$$

At the slave level, the adaptive path integral (see \cite{IA}),
representing an $\infty-$dimen-sional neural network,
corresponding to the adaptive bilateral joint action (\ref{Fey1}),
reads
\begin{equation}
\langle OUT|IN\rangle :=\put(0,0){\LARGE $\int$}\put(20,15){\small
$\Sigma$}\quad\,\mathcal{D}[w,x,y]\, {\mathrm e}^{\mathrm i
A[x,y;t_{i},t_{j}]}, \label{pathInt}
\end{equation}
where the Lebesgue integration is performed over all continuous paths $%
x^{i}=x^{i}(t_{i})$ and $y^{j}=y^{j}(t_{j})$, while summation is
performed over all associated discrete Markov fluctuations and
jumps. The symbolic
differential in the path integral (\ref{pathInt}) represents an {%
adaptive path measure}, defined as a weighted product
\begin{equation}
\mathcal{D}[w,x,y]=\lim_{N\rightarrow \infty
}\prod_{s=1}^{N}w_{ij}^{s}dx^{i}dy^{j},\qquad ({i,j=1,...,n}).
\label{prod}
\end{equation}

Similarly, in case of the triple joint action, the adaptive path
integral reads,
\begin{equation}
\langle OUT|IN\rangle :=\put(0,0){\LARGE $\int$}\put(20,15){\small
$\Sigma$}\quad\,\mathcal{D}[w;x,y,z;q]\, {\mathrm e}^{\mathrm i
A[t_{i},t_{j},t_{k};t]}, \label{pathInt2}
\end{equation}
with the adaptive path measure defined by%
\begin{equation}
\mathcal{D}[w;x,y,z;q]=\lim_{N\rightarrow \infty
}\prod_{S=1}^{N}w_{ijkr}^{S}dx^{i}dy^{j}dz^{k}dq^{r},\qquad
(i,j,k=1,...,n;~r=1,...,3n).  \label{prod2}
\end{equation}

The adaptive path integrals (\ref{pathInt}) and (\ref{pathInt2})
incorporate the {local Bernstein adaptation process}
\cite{Bernstein,Bernstein2} according to Bernstein's discriminator
concept
\begin{equation*}
desired\;state~SW(t+1)\;=\;current\;state~IW(t)\;+\;adjustment~step~\Delta
W(t).
\end{equation*}

\section{Discussion}

This paper has developed an adaptive path integral approach to
modelling topological phase transition, chaos and joint action in
the LSF--manifold. The traditional neural networks approaches are
known for their classes of functions they can
represent.\footnote{Here we are talking about functions in an
\emph{extensional} rather than merely \emph{intensional} sense;
that is, function can be read as input/output behavior
\cite{Barendregt,Benthem,Forster,Hankin}.} This limitation has
been attributed to their low-dimensionality (the largest neural
networks are limited to the order of $10^5$ dimensions
\cite{Izh}). The proposed path integral approach represents a new
family of function-representation methods, which potentially
offers a basis for a fundamentally more expansive solution.

This new family of function-representation methods is now capable
of representing input/output behavior of more than one actor.
However, as we add the second and subsequent actors to the model,
the requirements for the rigorous geometrical representations of
their respective LSFs become nontrivial. For a single actor or a
two--actor co--action the Riemannian geometry was sufficient, but
it becomes insufficient for modelling the $n$--actor (with $n\geq
3$) joint action, due to an {intrinsic chaotic coupling} between
the individual actors' LSFs. To model an $n$--actor joint LSF, we
have to use the Finsler geometry, which is a generalization of the
Riemannian one. This progression may seem trivial, both from
standard psychological point of view, and from computational point
of view, but it is not trivial from the geometrical perspective.

The robustness of biological motor control systems in handling
excess degrees of freedom has been attributed to a combination of
tight hierarchical central planning and multiple levels of sensory
feedback self--regulation that are relatively autonomous in their operation \cite%
{Bernstein3}. These two processes are connected through a
top--down process of action script delegation and bottom--up
emergency escalation mechanisms. There is a complex interplay
between the continuous sensory feedback and motion/action planning
to achieve effective operation in uncertain environments, such as
movement on uneven terrain cluttered with obstacles.

Complementing Bernstein's motor control principles is Brooks'
concept of computational {subsumption architectures}
\cite{BrooksLayered,BrooksElephants}, which provides a method for
structuring reactive systems from the bottom up using layered sets
of behaviors. Each layer implements a particular goal of the
agent, which subsumes that of the underlying layers. Similar
architectures have been proposed to account for the mechanism of
cognitive and motor working memory, sequence learning and
performance (see, e.g.,\cite{Gro08}). According to \cite%
{BrooksLayered,BrooksElephants}, a robot's lowest layer could be
``avoid an object", on top of it would be the layer ``wander
around", which in turn lies under ``explore the world". The top
layer in such a case could represent the ultimate goal of
``creating a map". In this configuration, the lowest layers can
work as reflexive mechanisms, while the higher layers contain
control logic implementing more abstract goals.

The substrate for this architecture comprises a network of finite
state machines augmented with timing elements. A subsumption
compiler compiles {augmented finite state machine} descriptions
into a special-purpose scheduler to simulate parallelism and a set
of finite state machine simulation routines. The resulting
networked behavior function can be described as:
\begin{equation*}
final\;state~w(t+1)\; =\; current\;state~w(t)\;+\;
adjustment~behavior ~f(\Delta w(t)).
\end{equation*}

The Bernstein {weights}, or {Brooks nodes}, $w^s_{ij}=w^s_{ij}(t)
$ in (\ref{prod}) are updated by the {Bernstein loop} during the
joint transition process, according to one of the two standard
neural learning schemes, in which the micro--time level is
traversed in discrete steps:
\begin{enumerate}
\item A {self--organized}, {unsupervised}  (e.g., Hebbian--like
\cite{Hebb}) learning rule:
\begin{equation}
w^s_{ij}(t+1)=w^s_{ij}(t)+ \frac{\sigma}{\eta}%
(w^{s,d}_{ij}(t)-w^{s,a}_{ij}(t)),  \label{Hebb}
\end{equation}
where $\sigma=\sigma(t),\,\eta=\eta(t)$ denote {signal} and {%
noise}, respectively, while new superscripts $d$ and $a$ denote {%
desired} and {achieved} micro--states, respectively; or

\item A certain form of a {supervised gradient descent learning}:
\begin{equation}
w^s_{ij}(t+1)\,=\,w^s_{ij}(t)-\eta \nabla J(t),  \label{gradient}
\end{equation}
where $\eta $ is a small constant, called the {step size}, or the
{learning rate,} and $\nabla J(n)$ denotes the gradient of the
`performance hyper--surface' at the $t-$th iteration,
\end{enumerate}
where $t=t_0,t_1,...,t_s$ then $t+1=t_1,t_2,...,t_{s+1}$.

Both Hebbian and supervised learning\footnote{%
Note that we could also use a reward--based, {reinforcement
learning} rule \cite{SB}, in which system learns its {optimal
policy}:
\begin{equation*}
innovation(t)=|reward(t)-penalty(t)|.
\end{equation*}%
} are used in local decision making processes, e.g., at the
intention formation phase (see \cite{IA}). Overall, the model
presents a set of formalisms to represent time-critical aspects of
collective performance in tactical teams. Its applications include
hypotheses generation for real and virtual experiments on team
performance, both in human teams (e.g., emergency crews) and
hybrid human-machine teams (e.g., human-robotic crews). It is of
particular value to the latter, as the increasing autonomy of
robotic platforms poses non-trivial challenges, not only for the
design of their operator interfaces, but also for the design of
the teams themselves and their concept of operations.

Some specific problems that Brooks poses in \cite{BrooksElephants}
include: (i) how to combine many (e.g., more than a dozen)
behavior generating modules in a way which lets them be productive
and cooperative; (ii) how to automate the building of interaction
interfaces between behavior generating modules, so that larger
(and hence more competent) systems can be built; and (iii) how to
automate the construction of individual behavior generating
modules, or even to automate their modification. An aggregation of
phase--transition--related {order parameters} within individual
actors need to be triggered (either by an internal or external
control mechanism) into collective alignment to be able to perform
the joint action. The `guiding' forces help to guide the better
alignment or fine--tunning of individual LSFs for useful
co--action outcomes to emerge. Using the combined LSF--geometry
approach proposed in this paper should provide some useful answers
to the above questions.

Here we remark that collective phase transitions, which have more
coupled degrees of freedom than individual ones, necessitate more
stringent constraints, to avoid/control a higher-dimensional
chaos. The sophisticated chaos--control techniques, including
constraining contextual boundaries -- choosing a target subspace
of the joint LSF-manifold -- and guiding force--fields, need to be
defined to alow desired collective behaviors to emerge from both
chaotic and non-chaotic sets of possible initial evolution
alternatives.

\section{Conclusion}

Extending the LSF model to incorporate the notions of
meta-stability, phase transitions and embedded geometrical chaos
has enabled representation of increased complexity in
goal-directed action, including the capability to represent joint
action by two or more co-actors. This capability remains
consistent with modern neuroscience theorizing linking
macro-behavioral meta-stability and phase transitions to
microscopic-level cortical neurodynamics.

There is a degree of correspondence between phase transition
mechanisms for cognitive performance in humans and transitions
between stable behavior states and competency levels in autonomous
robots. The approach developed in this paper offers a theoretical
framework to integrate observations and models of both individual
and collective robot behaviors and competencies, capable of coping
with the increased complexity of the real world. This framework
can both guide future substantive empirical work into collective
robot behaviors and be validated by it.

The new model developed in this paper offers substantial
improvements over the geometrical properties regarding
multiple-actor systems, due to the chaotic coupling between the
actors. We have also discussed how the proposed path integral
represents a new family of function representation techniques that
may expand on the range of function types that are currently
afforded by standard neural-network models. Specifically, we are
interested here in biologically plausible function representation
as a means for characterizing the input/output behavior of
multi-actor systems, and thus regard function representation in an
extensional sense. The full realisation of these possibilities in
practical applications is a subject of ongoing research.

\end{document}